\documentclass[aps,prd, preprintnumbers,groupedaddress,nofootinbib]{revtex4}
\pdfoutput=1
\usepackage{graphicx}
\usepackage{latexsym}
\usepackage{amsfonts}
\usepackage{amssymb}
\usepackage{amsmath}
\usepackage{slashed}
\usepackage{feynmp}
\usepackage{hyperref}
\usepackage{xspace}

\newcommand{\tev}{\ensuremath{\mathrm{\,Te\kern -0.1em V}}\xspace}
\newcommand{\gev}{\ensuremath{\mathrm{\,Ge\kern -0.1em V}}\xspace}
\newcommand{\tevt}{\ensuremath{\mathrm{Te\kern -0.1em V}}\xspace}

\begin{document}

\title{Stable Colored Particles R-SUSY Relics or Not?}

\author{Matthew R.~Buckley,$^{1,2}$ Bertrand Echenard,$^1$ Dilani Kahawala,$^3$ and Lisa Randall$^3$}
\affiliation{$^1$Center for Particle Astrophysics, Fermi National Accelerator Laboratory, Batavia, IL 60510}
\affiliation{$^2$Department of Physics, California Institute of Technology, Pasadena, CA 91125, USA}
\affiliation{$^3$Harvard University, Cambridge, MA 02138, USA}
\preprint{FERMILAB-PUB-10-310-T; CALT 68-2780}
\date{\today}

\begin{abstract}
R-hadrons are only one of many possible stable colored states that the LHC might produce. All such particles would provide a spectacular, if somewhat unusual, signal at ATLAS and CMS. Produced in large numbers and leaving a characteristic signature throughout all layers of the detector, including the muon chamber, they could be straightforward to discover even with low luminosity. Though such long lived colored particles (LLCPs) can be realized in many extensions of the Standard Model, most analyses of their phenomenology have focused only on R-hadrons. In order to distinguish among the possibilities, fundamental quantum numbers of the new states must be measured. In this paper, we demonstrate how to identify the $SU(3)_C$ charge and spin of such new particles at the LHC.
\end{abstract}

%\pacs{FERMILAB-PUB-10-310-T; CALT 68-2780}

%%%%%%%%%%%%%%%%%%%%%%%
\maketitle

\section{Introduction \label{sec:intro}}

After decades of anticipation and preparation, the Large Hadron Collider will shortly open the door to $\tevt$-scale physics. This energy range is of great theoretical interest, as it has long been suspected of holding the answers to electroweak symmetry breaking, and the associated naturalness problem \cite{Weinberg:1975gm,Weinberg:1979bn,Susskind:1978ms,t-Hooft:1979qy}. From technicolor \cite{Weinberg:1979bn,Susskind:1978ms,Weinstein:1973gj}, to supersymmetry \cite{Wess:1974tw} and extra dimensions 
\cite{Antoniadis:1990ew,Antoniadis:1993jp,ArkaniHamed:1998nn,ArkaniHamed:1998rs,Antoniadis:1998ig,Randall:1999ee,Randall:1999vf} a great deal of effort has gone into discovering possible solutions to these problems, and determining the associated collider signatures. However, we don't yet know what will appear at the weak scale and we want to be open to the broadest range of possibilities.  It is critical to also consider the experimental signatures of other scenarios for new TeV-scale physics, ones that may not easily fit into the known solutions for the various problems of the Standard Model (SM). 

In this paper, we propose methods to measure both the spin and $SU(3)_C$ color charge of strongly interacting massive particles that are stable on detector timescales. Though most of the detailed analyses have focused on (meta-)stable gluinos or squarks, supersymmetric $R$-hadrons are just one realization of strongly interacting, stable particle. We will take the most general possible viewpoint, and ask simply about the quantum numbers of the colored state, independent of the model in which it might originate. Examples abound in the literature, including universal extra dimensions \cite{Appelquist:2000nn,Appelquist:2002wb,Cheng:2002iz,Macesanu:2002db} that can mimic many features of SUSY models; unusual spectra, such as charged lightest KK-odd modes \cite{Feng:2003nr,Shah:2006gs}, are also possible and may be a strongly interacting state. More exotic models have also been proposed that would include (meta-)stable colored particles: KK-towers of $X$ and $Y$ grand unified gauge bosons in warped extra-dimensions with GUT-parity \cite{Goldberger:2002pc,Nomura:2003qb,Nomura:2004is,Nomura:2004it}, long-lived leptoquarks \cite{Friberg:1997nn}, $4^{\rm th}$ generation quarks \cite{Fishbane:1983hf,Fishbane:1984zv}, mirror fermions \cite{Barbieri:2005ri,He:2001tp}, perhaps in vector-like generations \cite{Banks:1986cg,Polonsky:2000zt}, or related to symmetries stabilizing the dark sector \cite{Walker:2009ei}. For a larger list of possible models and particle candidates, see Ref.~\cite{Fairbairn:2006gg}. In this paper, we will use ``long-lived-colored particle" or `LLCP' as a generic name for any new stable colored particle. 

All these models generate similar signatures in the detector. Many are produced with very large cross-sections, making discovery in early running a possibility. As they are both strongly interacting and stable, they will pass through the entire detector. If the particle hadronizes into charged states, it will deposit energy in the central tracker, electronic and hadronic calorimeters, and be visible in the muon chambers \cite{Drees:1990yw,Fairbairn:2006gg}. Thus, such particles will present a striking signature at the LHC, initially appearing as ``heavy muons'' in events with no missing $p_T$ (assuming both LLCPs hadronize into charged objects) that would be extremely difficult to replicate by a SM background. Additionally, as will be discussed in greater detail, the LLCPs often undergo nuclear interactions in the detector which rehadronize the particle and allow for the charge to switch sign. This can result in another unique signature, though specialized tracking procedures may be necessary to take full advantage of this. Finally, an alternative search strategy is to look for stopped tracks in the detector volume \cite{Arvanitaki:2005nq}. Such searches have been carried out at D0 \cite{Abazov:2007ht}, CMS \cite{Kazana:2009zz}, and ATLAS \cite{atlassearch}.

With discovery a relatively straightforward issue, in this paper we concern ourselves with the problem of identifying the underlying quantum numbers of the new state. If we are to determine whether a stable $SU(3)_C$-charged particle is truly a gluino, a squark, a UED gluon KK=1 mode, or some other expression of new physics, it will be necessary to measure the LLCP mass, spin, and charge under the SM gauge groups. 

Of these, mass is a straightforward measurement: time of flight information will be sufficient to determine the mass to good accuracy \cite{Kilian:2004uj}. In this work we demonstrate techniques for measuring both the spin and $SU(3)_C$ charge of LLCPs. In Section~\ref{sec:spin}, we demonstrate the former measurement; we will show that spin can be determined from the polar angle differential cross-section in LLCP pair-production events. Unlike most proposed new physics events, pair production of LLCPs have almost no missing energy, so this distribution can be reconstructed without ambiguity. Even with the presence of $t$-channel diagrams, which cause forward peaks in the distribution for all possible spin assignments, sufficient differences remain in the distributions, which allow identification of this critical quantity.

In Section~\ref{sec:color}, we demonstrate techniques to identify, with some limitations, the color charge of new stable particles. In particular, we show that it is possible to distinguish the production of a stable pair of particles in octet representations of $SU(3)_C$ ({\it e.g.}~gluinos) from production of particle-antiparticle pair in triplet/anti-triplet representations ({\it e.g.~} stops). This method relies on the fundamental asymmetry present in the detectors: they are built from baryons, rather than anti-baryons. As a result, the hadronization of a triplet of $SU(3)_C$ follows a very different path from that of an anti-triplet, leading to a measurable difference in energy deposition. 

Perhaps the best known realization of such particles is in supersymmetry, where in some schemes gluinos or squarks can be the lightest (or next to lightest) supersymmetric particle \cite{Baer:1998pg,Mafi:1999dg,Mafi:2000kg,Raby:1997bpa,Raby:1997pb}. In this case, the strongly interacting particles are stabilized by an unbroken (or weakly broken if the particles are only meta-stable \cite{Dreiner:1997uz,Berger:2003kc}), $R$-symmetry. As such, they have become known as $R$-hadrons \cite{Farrar:1978rk,Farrar:1978xj}. Searches for such particles have been performed at ALEPH \cite{Heister:2002hp}, CDF \cite{Acosta:2003ys}, and LEP2 \cite{SUSYwg}, and exclude particles with mass less than about $200-250 \gev$, depending on the theoretical assumptions made. Searches are planned at both ATLAS \cite{Mermod:2009ct} and CMS.

The physics in the early Universe may provide significant constraints if these strongly interacting particles are truly stable (or at least have a lifetime much longer than the age of the Universe). Both direct searches for dark matter %\cite
and searches for anomalously heavy seawater \cite{Smith:1982qu} preclude dark matter from having $SU(3)_C$ charge. 
This places strong limits on the mass of any new stable colored particle; gluinos, for example, can evade cosmological bounds only if their masses are less than about a $\tevt$ \cite{Baer:1998pg}, and seawater tests may lower the allowed mass to $\sim100 \gev$.

Of course, at the LHC, a particle needs only live longer than a few dozen nanoseconds to be seen as `stable.' In this case, the  constraints are relaxed and depend on  lifetime. Again specializing to the case of long-lived gluinos, Ref.~\cite{Arvanitaki:2005fa} finds that the SM's successful prediction of nuclear abundances from Big Bang Nucleosynthesis excludes lifetimes greater than $100~$seconds (see Ref.~\cite{Kawasaki:2004qu} for a more-in-depth discussion of hadronic decays in this epoch). Lifetimes up to $10^{13}$~seconds are excluded, as they would distort the cosmic microwave background, while lifetimes on the order of the age of the Universe are ruled out by observations of the diffuse gamma ray background by EGRET \cite{Sreekumar:1997un}. From this, we conclude that any new colored particles at the LHC must either decay within 100~seconds, or have a lifetime significantly longer than the age of the Universe. We consider such possibilities below.

\section{LLCP Spin Measurements \label{sec:spin}}

Measuring the spin of new particles at the $\tevt$-scale has long been recognized as a critically important task in identifying the underlying theory. While the total cross section may be used as a spin measurement, we are interested in a more reliable and less indirect method. Techniques developed for supersymmetric particles or similar physics ({\it e.g.}~\cite{Barr:2004ze,Barr:2005dz,Battaglia:2005zf,Smillie:2005ar,Wang:2006hk,Alves:2007xt,Buckley:2007th,Buckley:2008eb,Buckley:2008pp,Murayama:2009jz}), are not applicable to stable particles. However, we can rely on simpler methods, since the event is fully reconstructible. In particular, measurement of the angular distribution via the differential cross-section with respect to the polar angle $\theta^*$ in the center of mass (c.o.m.) frame is sufficient to determine the spin of pair-produced particles. Though the presence in some models of $t$-channel production tends to produce forward peaks at large values of $|\cos\theta^*|$, enough information remains to make spin measurement possible.

We consider several possible cases: the production of massive triplet/anti-triplet fermions and scalars, as well as octet vectors. The minimal models add only the LLCPs themselves, in which case new physics Lagrangians are just
\begin{eqnarray}
{\cal L}_{\rm scalar} & = &(D_\mu Q_S) (D^\mu Q_S)^\ast - M^2 Q_S^\ast Q_S \label{eq:scalarL} \\
{\cal L}_{\rm spinor} & = & i\bar{Q}_F \slashed{D} Q_F - M \bar{Q}_FQ_F  \label{eq:spinorL} \\
{\cal L}_{\rm vector} & = & G_{\mu\nu}G^{\mu \nu} - M^2 Q_{V,\mu} Q_V^\mu \label{eq:vectorL}
\end{eqnarray}
An example of $Q_S$ includes a supersymmetric quark, while the fermion $Q_F$ can be a $4^{\rm th}$ generation quark, but more generally any triplet fermion representation of Standard Model quantum mnumbers.  The octet vectors are realized as $KK=1$ gluons in UED, though here we have integrated out the $KK=1$ quarks.

However, in most complete extensions of the SM that contain potential LLCPs, additional new states that can couple to the LLCPs, quarks and/or gluons. We therefore include the addfitional cases of up-type squark $R$-hadrons with gluino intermediaries, UED up-type $KK=1$ quarks with heavier $KK=1$ gluon intermediaries, and $KK=1$ gluons with heavier up-type $KK=1$ quark intermediaries.

In all of the models, presence of $t$-channel diagrams create forward peaks in the $|\cos\theta^*|$ distribution. It is generally held that such distortions make spin determination difficult (see, for example \cite{Battaglia:2005zf}). However, while the scalar and spinor distributions do develop similar peaks at large values of $|\cos\theta^*|$, we show that enough qualitative differences remain to distinguish the various scenarios \cite{Allanach:2001sd}.  However,  particular choices of intermediary masses can confuse the issue and make the differential cross sections appear to be degenerate.

\begin{figure}[ht]
\includegraphics[width=0.55\columnwidth]{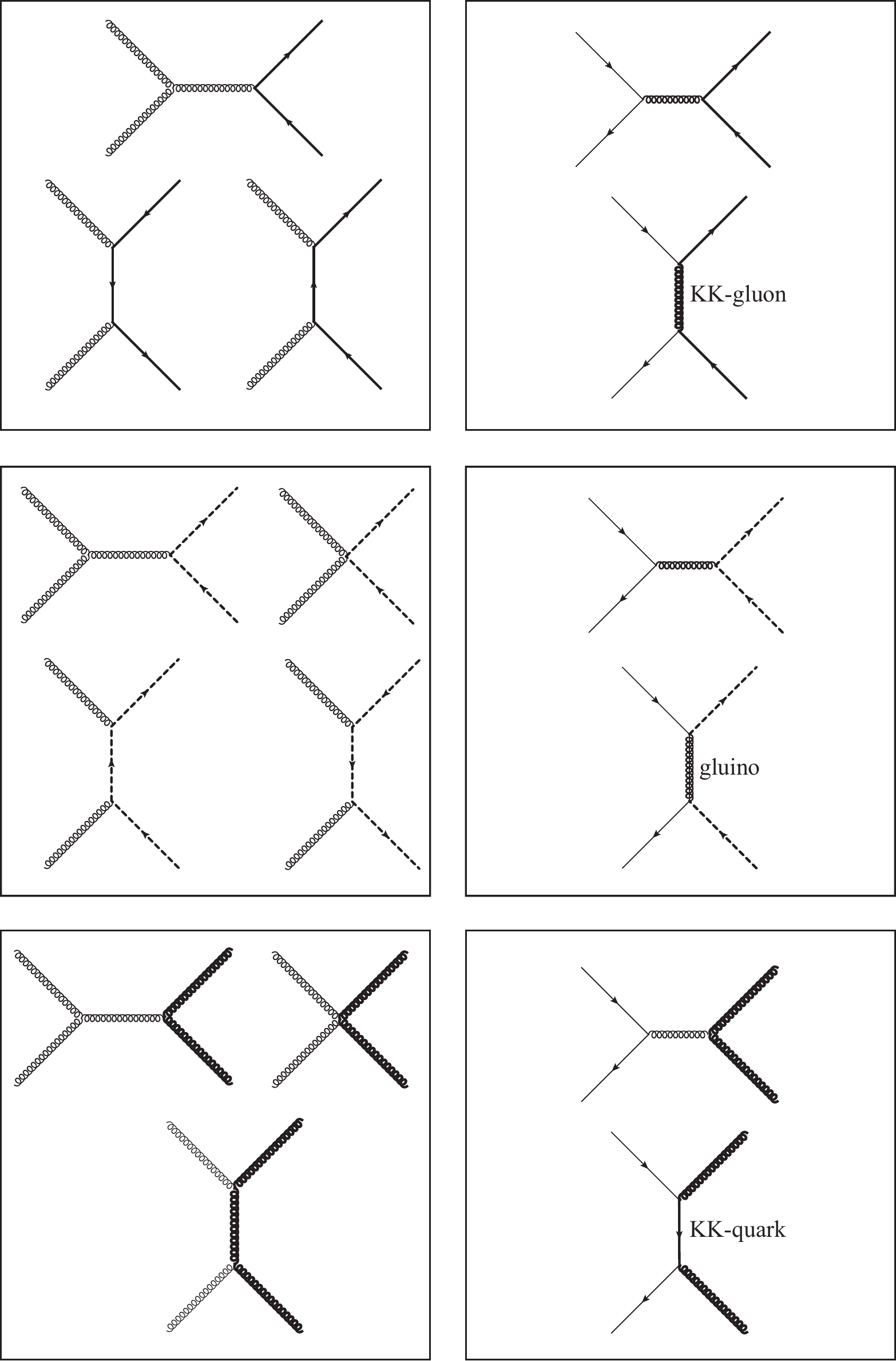}
\caption{Feynman diagrams for the production of spinor (top), scalar (middle), and vector (bottom) LLCPs. Left panels show the diagrams proceeding from gluon initial states, while the quark-antiquark diagrams are on the right. The diagrams requiring the presence of additional heavy states (KK-gluons, -quarks, or gluinos) are labeled.  \label{fig:feynman}}
\end{figure}

The analytic formula for the pair production in proton-proton collisions in each case are straightforward to derive. The relevant Feynman diagrams for quark-antiquark and gluon initial states are shown in Fig.~\ref{fig:feynman}. For each model, the differential cross-section is convolved with the parton distribution functions (p.d.f.s) using the CTEQ5 p.d.f.~\cite{Kuhlmann:1999tw}.

\begin{figure}[ht]
\includegraphics[width=0.45\columnwidth]{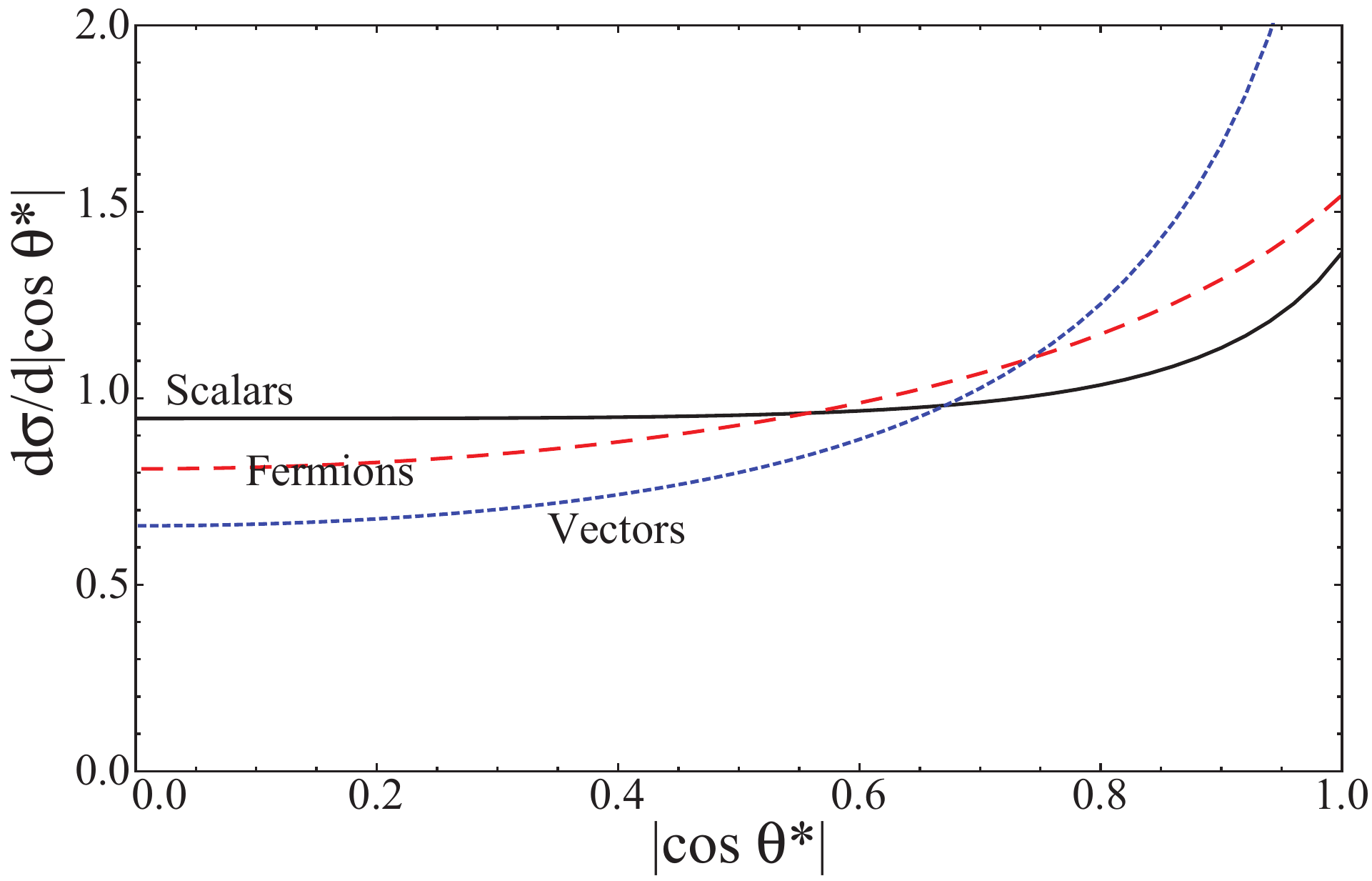}\includegraphics[width=0.45\columnwidth]{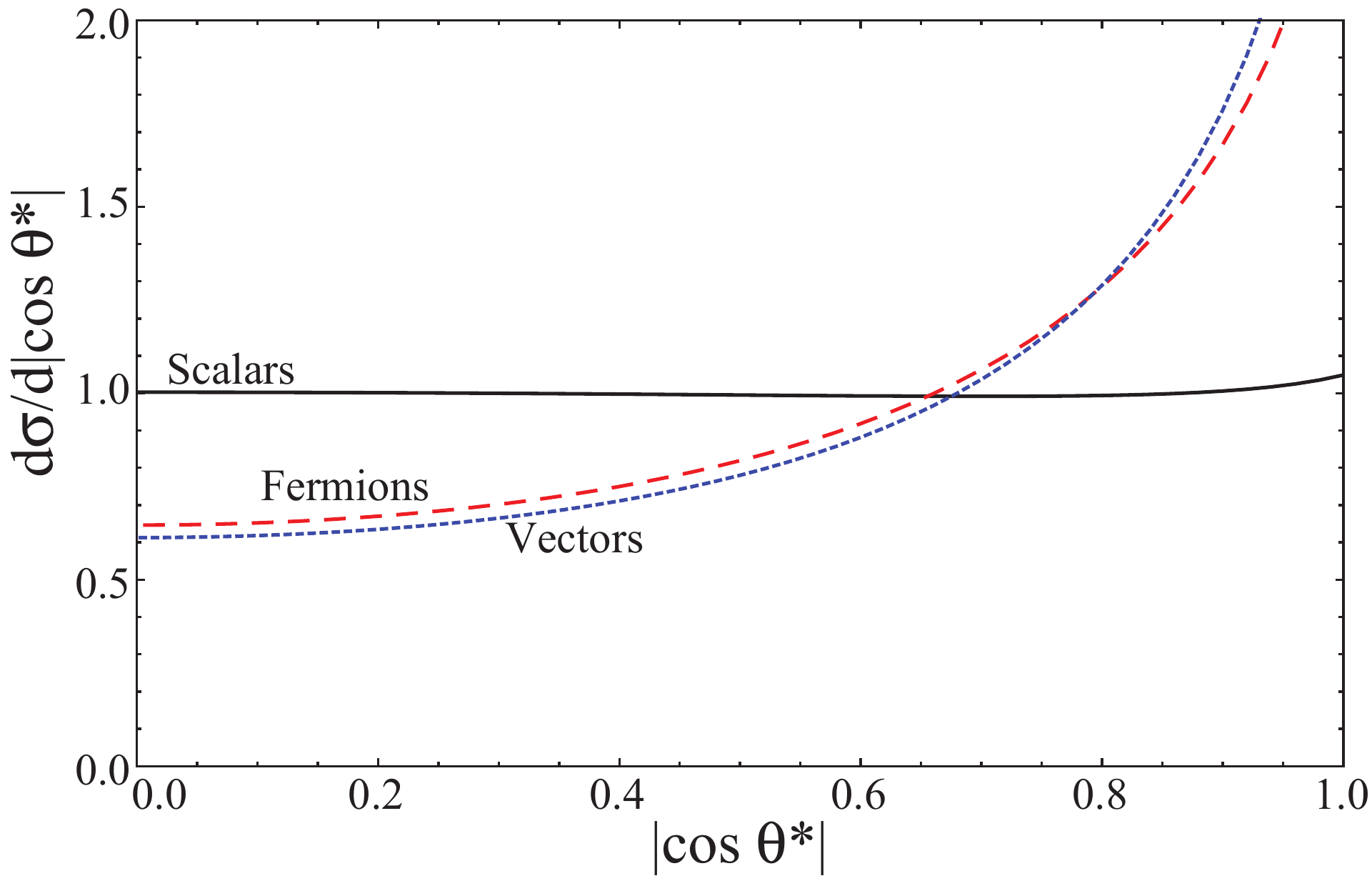}
\includegraphics[width=0.45\columnwidth]{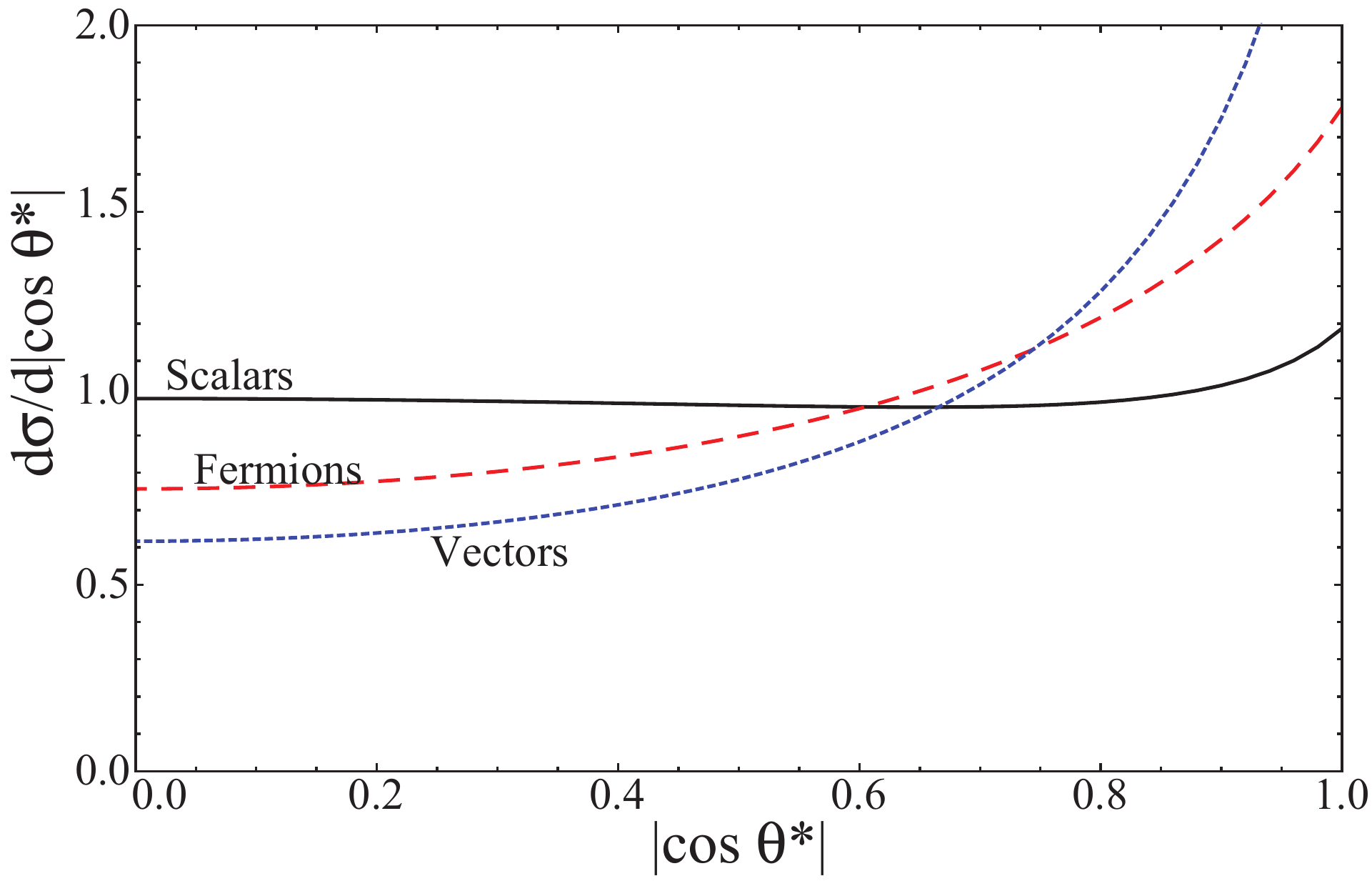}
\caption{The normalized differential cross-sections $\sigma^{-1}d\sigma/d|\cos\theta^*|$ for pair production of $500 \gev$ LLCPs in $pp$ collisions at $\sqrt{s}=10 \tev$. No cut on the pseudo-rapidity or velocity $\beta$ of each LLCP is applied. Left: minimal scalars, fermions and vectors, as introduced in Eqs.~(\ref{eq:scalarL})-(\ref{eq:vectorL}). Right and Center: up-type squarks with gluino intermediaries, up-type $KK=1$ quarks with $KK=1$ gluon intermediaries. The intermediary mass is $700 \gev$ for the upper right, and $1000\gev$ for the lower center plot.  \label{fig:nocutdist}}
\end{figure}

In Fig.~\ref{fig:nocutdist}, we show the differential cross-sections after convolution before any acceptance cuts assuming a LHC center of mass energy of $\sqrt{s}=10 \tev$ and a LLCP mass of $500 \gev$.  For models that have a heavy intermediary ({\it i.e.}~squarks with a heavy gluino, KK-quarks with a heavy KK-gluon, and KK-gluons with a heavy KK-quark), we choose two masses of the heavy state: $700 \gev$ and $1000\gev$. When no intermediaries are present (or are very heavy), all three spin assignments have significantly different differential cross sections, and so can be distinguished with relative ease. However, in the case of $700\gev$ intermediaries, the differential cross sections of fermions and vectors are similar, making discrimination very difficult. In all cases, the cross-sections is normalized to 1.

\begin{table}[ht]
\begin{tabular}{|c||c|}
\hline 
Model &  Cross-section (fb) after cuts \\ \hline \hline 
Minimal scalars & 18 \\ \hline
Minimal spinors & 130 \\ \hline
Minimal vectors &  $1.3 \times 10^4$ \\ \hline
Up squarks with $700\gev$ gluinos & 29 \\ \hline
$KK=1$ up quarks with $700\gev$ $KK=1$ gluons & 340 \\ \hline
$KK=1$ gluons with $700\gev$ $KK=1$ quarks & $1.2 \times 10^4$ \\ \hline
Up squarks with $1000\gev$ gluinos & 24 \\ \hline
$KK=1$ up quarks with $1000\gev$ $KK=1$ gluons & 210 \\ \hline
$KK=1$ gluons with $1000\gev$ $KK=1$ quarks & $1.3 \times 10^4$ \\ \hline
\end{tabular}
\caption{Total cross-section assuming $\sqrt{s}=10 \tev$ and LLCP mass of $500 \gev$ after $|\eta|<2.1$ and $\beta>0.6$ cuts. Heavy intermediary particles are chosen to be $700 \gev$ or $1000\gev$ (see text). \label{tab:sigma}}
\end{table}

We next impose the cut $|\eta|< 2.1$ to ensure that both LLCPs end up inside the barrel regions of the ATLAS and CMS detectors, and a cut of $\beta > 0.6$, which is necessary for the heavy muon trigger to identify the correct bunch crossing \cite{Mermod:2009ct}. Although these cuts tend to remove events at large $|\cos\theta^*|$ ($t$-channel production diagrams generate forward peaks close to the beam-line) they do not greatly affect our ability to discriminate spin, as the differential cross-section at small values of $|\cos\theta^*|$ has more resolving power. It should be noted that future work by the experiments on the ``heavy muon'' triggers may allow the $\eta$ acceptance to be increased, perhaps up to $|\eta|< 2.5$. 

Note that the $|\cos\theta^*|$ distribution itself will not be affected by hadronization, as this energy scale $\sim \Lambda_{\rm QCD}$ is much less than the momenta of the particles themselves ($\sim 100\gev$). As we are considering exclusive LLCP pair production, our sample does not contain additional hard jets -- due to radiation of high-$p_T$ gluons for example -- which would have sufficient energy to significantly affect the differential cross section.

\begin{figure}[ht]
\includegraphics[width=0.45\columnwidth]{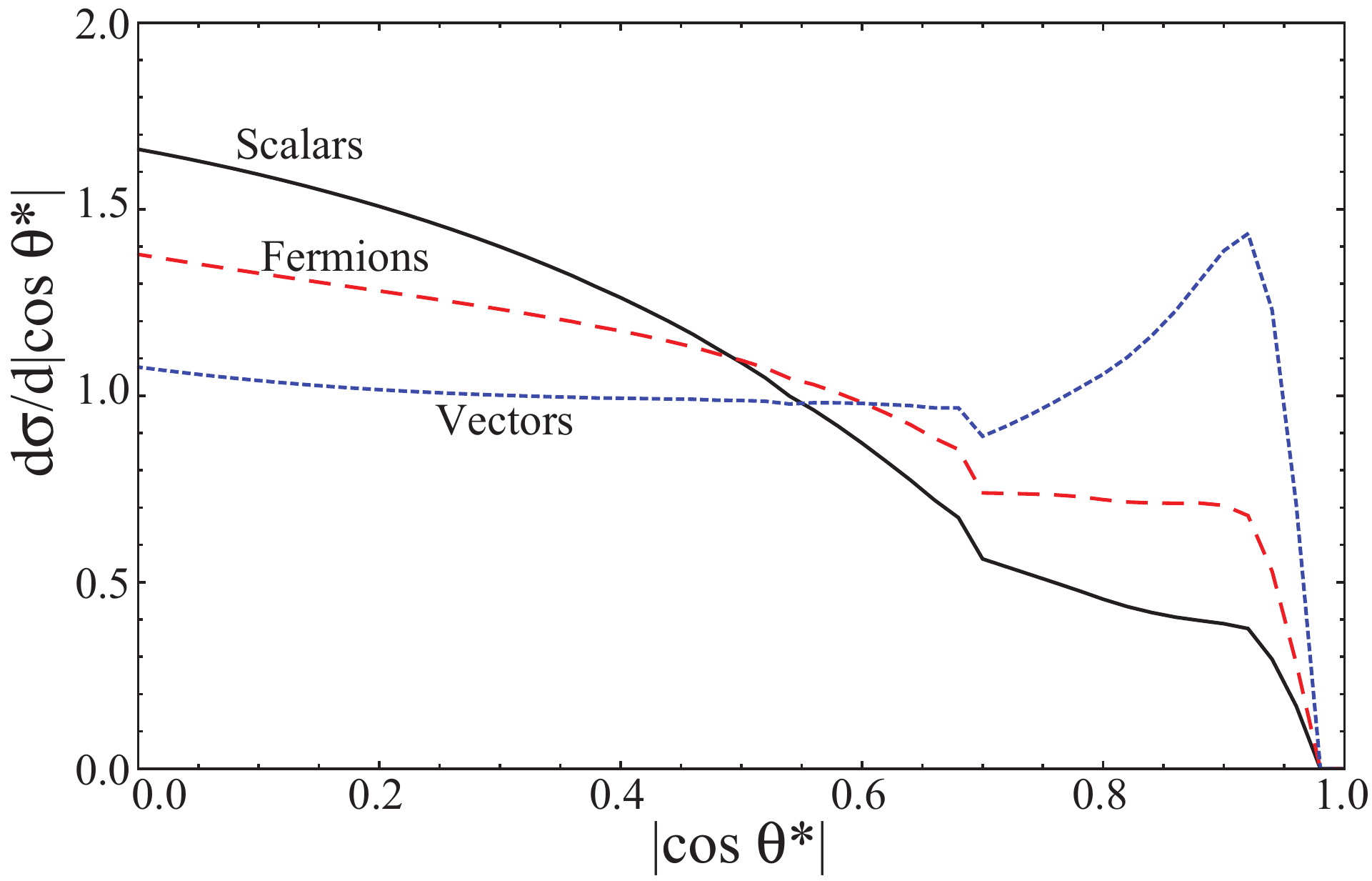}\includegraphics[width=0.45\columnwidth]{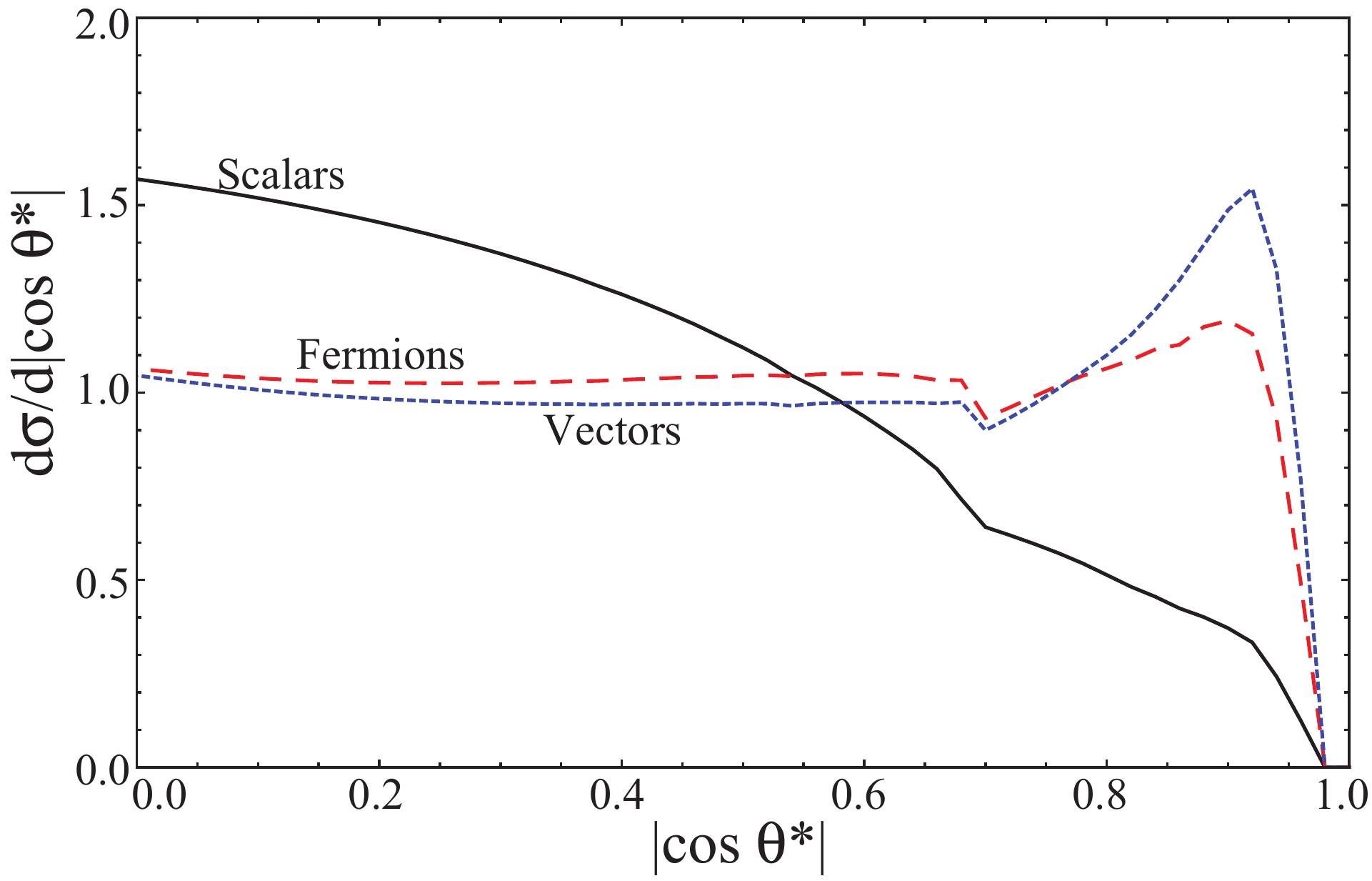}

\includegraphics[width=0.45\columnwidth]{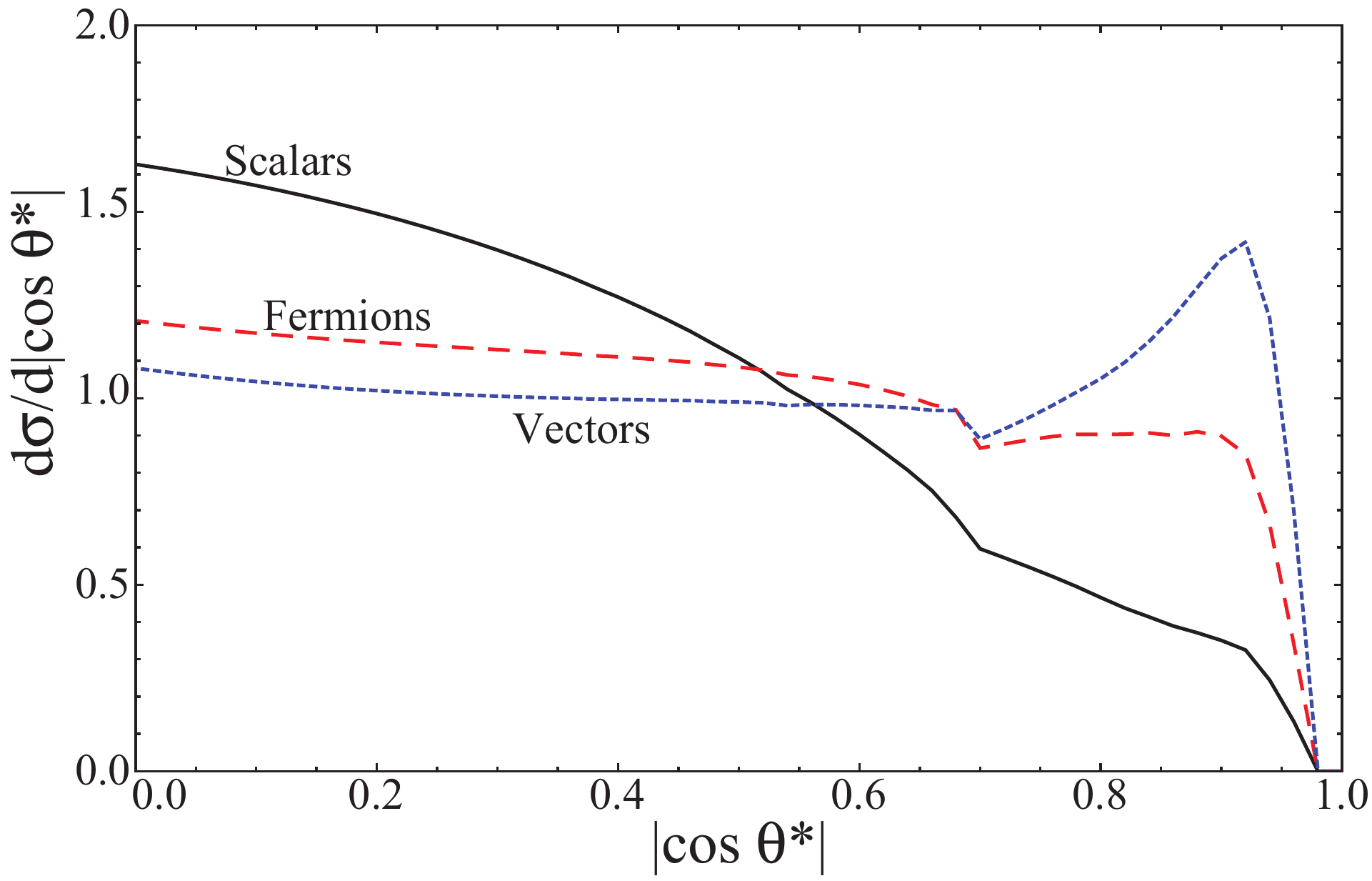}
\caption{The normalized differential cross-section $\sigma^{-1}d\sigma/d|\cos\theta^*|$ for pair production of $500 \gev$ LLCPs in $pp$ collisions at $\sqrt{s}=10 \tev$, requiring that both LLCPs are produced in the pseudo-rapidity range $|\eta| <2.1$ and have $\beta > 0.6$. Left: minimal scalars, fermions and vectors, as introduced in Eqs.~(\ref{eq:scalarL})-(\ref{eq:vectorL}). Right and Center: up-type squarks with gluino intermediaries, up-type $KK=1$ quarks with $KK=1$ gluon intermediaries. The intermediary mass is $700 \gev$ for the upper right, and $1000\gev$ for the lower center plot.  Numerical instabilities from the application of the c.o.m. $\eta$ cuts distort the curves around $|\cos\theta^*|=0.7$. \label{fig:dist}}
\end{figure}

Requiring both LLCPs to be produced with $|\eta|<2.1$ and $\beta>0.6$, we present the resulting total cross sections in Table~\ref{tab:sigma} and the differential cross sections are displayed in Fig.~\ref{fig:dist}. In most cases, the various models have significantly different distributions. We note that if we do a more model-dependent analysis and allow intermediate states of varying mass,  the fermion and vector cases will be degenerate for certain parameter choices. Of course, as the intermediary mass increases, the spectrum will revert to the `minimal' case, where the differential cross sections differ significantly. We estimate that distinguishing these differential cross sections may require $\sim 5$ bins in $|\cos\theta^*|$ with $\sim 1000$ events per bin. Lacking a full detector simulation, we estimate the efficiencies for production and detection of a charged-LLCP pair as ${\cal O}(0.1)$. Combined with our assumption of $5000$ binned events, the production cross-sections (which are very large in the fermion and vector cases, Table~\ref{tab:sigma}) imply that a spin measurement should be possible with integrated luminosity of ${\cal O}(1 \mbox{fb}^{-1})$ for the vector case, ${\cal O}(100 \mbox{fb}^{-1})$ for the fermions, and ${\cal O}(1000 \mbox{fb}^{-1})$ in the scalar case. 

\section{LLCP Color Charge Measurement \label{sec:color}}

In addition to spin determination, we would want to know the color charge of LLCPs as this provides key insights into the particle's identity and the associated underlying theory. Although some information could in principle be determined by measuring the total cross-section, we are again interested in a more direct handle on this quantum number. In this section, we  demonstrate a method to distinguish particles in a triplet/anti-triplet ($\bf{3} / \bar{\bf{3}}$) representation of $SU(3)_C$ from particles in an octet ($\bf{8})$. Further work is required to extend this method to representations other than the fundamental and adjoint.

The key element of this technique is the inherent asymmetry of detectors, built as they are of matter rather than antimatter. During its transit of the detector, an LLCP will undergo several nuclear interactions with the detector material, each of which has a significant probability of causing the LLCP to rehadronize by exchanging light colored particles with the nucleon \cite{Milstead:2009qy}. This introduces an asymmetry between LLCPs in a $\bf{3}$ representation versus ones in $\bar{\bf{3}}$: the former is interacting with many particles in the same representation as itself, while the latter sees essentially no light anti-quarks with which to hadronize. That means that after passing through the experiment the final mix of hadronized states for triplet states would significantly differ from that of antitriplets. As we shall show, the preferred state of triplet LLCPs  is an LLCP-baryon, that of an antitriplet LLCP is a meson.

The scattering of an LLCP with matter proceeds through the interaction of the light quark/gluon content with the target nucleus, as the probability of interaction between a heavy parton (the LLCP) and a quark at rest is proportional to the inverse square of the parton mass. In this context, the massive particle can be pictured as a stable non-interacting heavy parton, surrounded by a cloud of light quarks/gluons that scatter with the detector material. The cloud carries only a fraction of the total energy, and the mass of the nucleon is comparable to the total energy in the center of mass frame for the scattering. Interactions of LLCP-mesons that undergo a baryon number exchange (ending with the proton or neutron being destroyed) are kinematically favored over events that do not have such an exchange \cite{Kraan:2004tz}. Briefly, this is because the rest mass of the nucleon is about the same as the total available energy in the scattering. As a result, having a nucleon in the final state consumes nearly all of the available energy. For example, the phase space for a LLCP-meson + nucleon scattering to go into a LLCP-baryon + pions is much larger than for a LLCP-meson to LLCP-meson event.  An LLCP-meson will therefore preferentially undergo a baryon-exchange scattering with a nucleon, resulting in an LLCP-baryon and a shower of light mesons.

Once an LLCP-baryon is produced, the phase space to scatter back into an LLCP-meson is very small, as this requires the creation of a SM baryon which is heavy compared to the available energy in the scattering. On the other hand, an anti-triplet meson cannot undergo (anti-)baryon exchange to convert into an LLCP-anti-baryon; and if hadronized as a $\overline{(LLCP)}_3\bar{q}\bar{q}$, the preferred scattering is into a $\overline{(LLCP)}_3 q$-meson, destroying the nucleon in the process. This result is fairly robust and depends purely on phase space arguments and the relatively small mass splitting between LLCP-baryons and LLCP-mesons. 

In a similar fashion to the re-hadronization process, energy deposition in the detector differs between triplet and anti-triplet LLCPs. In a greatly simplified model, the ``black disk approximation,'' each light quark or gluon in the bound state contributes 12~mb to the nuclear scattering cross-section \cite{Mackeprang:2006gx}. Ignoring electromagnetic interactions, the LLCP triplet $(LLCP)_3$, hadronized as it is with two light quarks, on average scatters twice as often as the $\overline{(LLCP)}_3$-meson, and thus deposits twice as much energy. On the other hand, both octet LLCPs are produced in the same representation and so a pair of them will leave, on average, equal amounts of energy. Assuming that LLCPs will be pair produced at the LHC (since this is typically the case in theories containing such particles, this assumption is not overly restrictive), one can straightforwardly probe the color quantum number of the LLCP by looking for an asymmetry between energy deposition of the two tracks in the hadronic calorimeter. 

The black disk approximation is useful for illustrative purposes, but is obviously insufficient for detailed calculations. In Ref.~\cite{deBoer:2007ii}, a more sophisticated scattering model based on Regge phenomenology and low-energy hadron-hadron data was developed. As expected, the LLCP-baryon scattering cross-section is about twice as large as that of LLCP-mesons, owning to the additional light quark. The cross-section of LLCP-anti-baryons, due to a dominant annihilation process with baryons at low energies, is also larger than that of LLCP-baryons. Similarly, the $(LLCP)_3$-meson have a larger cross-section than  $\overline{(LLCP)}_3$-meson, since baryon-exchange processes are only permitted for LLCPs containing light antiquarks. The $(LLCP)_8$-meson and $(LLCP)_8$-gluon cross-sections are taken to be the sum of the $(LLCP)_3$ and $\overline{(LLCP)}_3$-meson, while that of $(LLCP)_8$-baryon is 50\% larger than the corresponding $(LLCP)_3$-baryon cross-section. We use the GEANT4 \cite{Agostinelli:2002hh} implementation of this model as described in Refs.~\cite{Milstead:2009qy,Mackeprang:2009ad}. This also includes electromagnetic energy losses through ionization, in addition to energy loss through nuclear scattering. As both LLCP-baryons and LLCP-mesons can be electrically charged, we expect that the presence of electromagnetic deposits will serve to shift both energy deposition curves to higher values. 

We illustrate our idea using particles in the triplet/anti-triplet representation with charge $\pm 2/3$ ({\it e.g.}~top squarks: $(LLCP)_3 = \tilde{t}, \overline{(LLCP)}_3=\bar{\tilde{t}}$) and neutral particles in the octet representation ({\it e.g.}~gluinos: $(LLCP)_8 = \tilde{g}$). The charged triplets can form LLCP-mesons with charge $+1$  ($(LLCP)_3 \bar{d}$), zero ($(LLCP)_3\bar{u}$ and $\overline{(LLCP)}_3u$) or $-1$ ($\overline{(LLCP)}_3d$), as well as charged LLCP-baryons (LLCP-anti-baryons), the lightest being $(LLCP)_3ud$ ($\overline{(LLCP)}_3\bar{u}\bar{d}$). Higher spin LLCP-baryons (LLCP-anti-baryons) are expected to decay to the ground state before interacting with the detector. The mass spectrum adopted is similar to the one used in Refs.~\cite{Mackeprang:2009zs,Kraan:2004tz}, in which the lightest LLCP-baryon is $\sim 0.3 \gev$ heavier than the massive particle, and the lightest LLCP-meson is $\sim 0.7 \gev$ heavier. These results are consistent with calculations using different approaches \cite{Chanowitz:1983ci,Buccella:1985cs,Foster:1997nm,Foster:1998wu,Gates:1999ei}. The two neutral mesons may allow the triplet LLCP-hadron to mix into the anti-triplet. This might occur via chargino/$W$ exchange in SUSY models. Since the level of mixing is model dependent, we consider two limiting cases: no mixing and maximal mixing, in which a neutral state has a 50\% probability in oscillating to its anti-particle. This corresponds to infinite and zero oscillation lengths, respectively. The lightest hadrons formed by the neutral octet include LLCP-mesons with charge $+1$  ($(LLCP)_8 u\bar{d}$), zero ($(LLCP)_8 q\bar{q}$ with $q=u,d$) or $-1$ ($(LLCP)_8 \bar{u}d$), the LLCP-glueball ($(LLCP)_8 g$) and the LLCP-baryon ($(LLCP)_8 uds$). Although their spectrum is not as well understood as the $(LLCP)_3$ examples, it is expected that the $(LLCP)_8$-mesons ($(LLCP)_8q\bar{q}, q=u,d$, $(LLCP)_8u\bar{d}$, {\it etc}) will be closely degenerate, and similar in mass to the lightest LLCP-baryon: $(LLCP)_8uds$ (see Ref.~\cite{Mackeprang:2009ad} and references therein).

\begin{figure}[th]

\includegraphics[width=0.45\columnwidth]{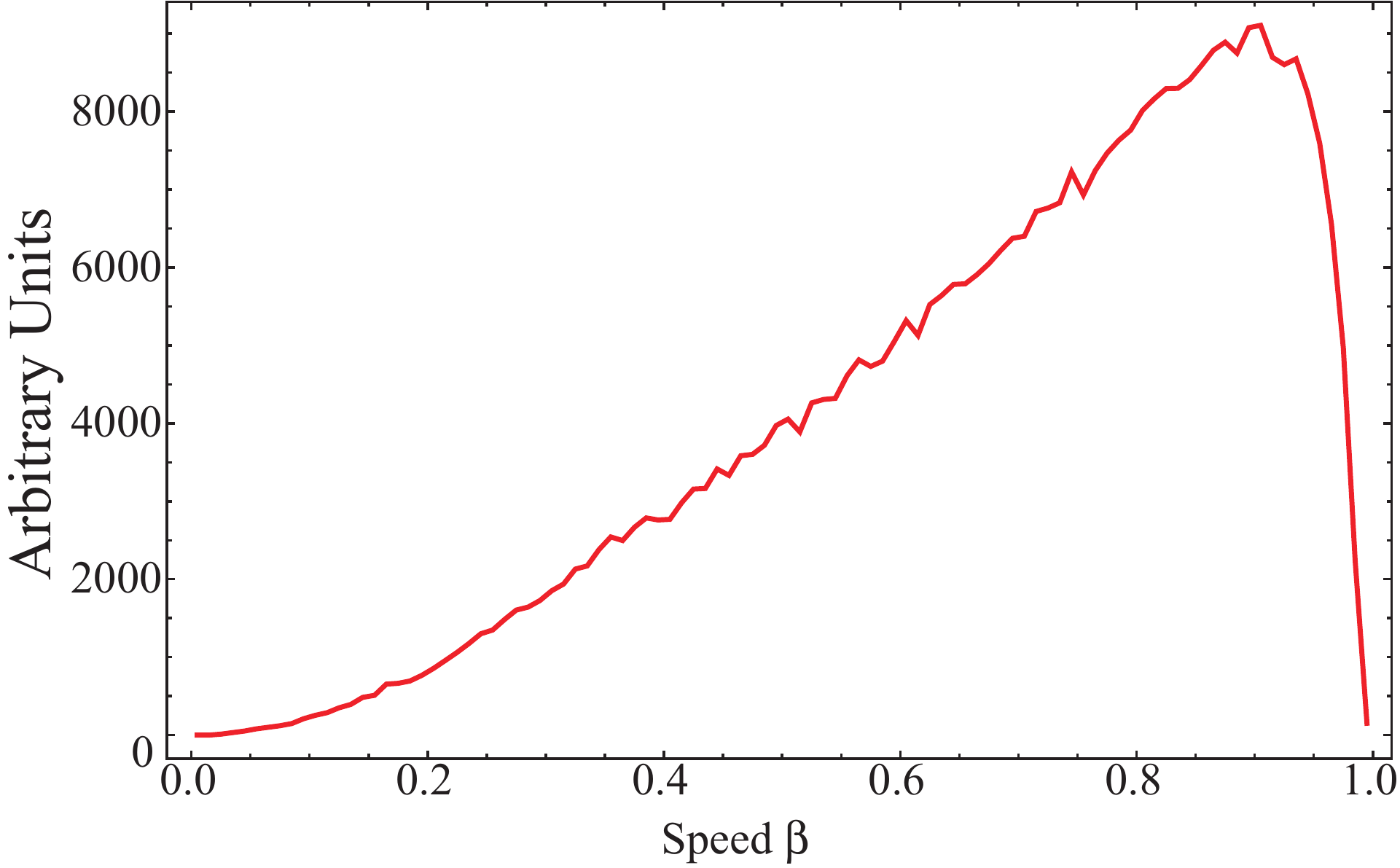}
\caption{The initial velocity distributions of the LLCPs. These distributions are obtained from MadGraph \cite{Alwall:2007st} simulations of pair-produced $500 \gev$ particles at the LHC with $\sqrt{s}=10 \tev$. This quantity depends mainly on the production kinematics, with only minor differences between the various spin models considered in Section~\ref{sec:spin}. We require $\beta>0.6$ to simulate the heavy muon trigger as planned by ATLAS \cite{Mermod:2009ct}. \label{fig:initialenergy}}
\end{figure}

The passage of LLCPs through matter is analyzed by firing LLCP beams initially composed of $100\%$ of either $(LLCP)_3\bar{d}$-mesons, $\overline{(LLCP)}_3d$-mesons, $(LLCP)_3ud$-baryons, $\overline{(LLCP)}_3\bar{u}\bar{d}$-anti-baryons, $(LLCP)_8u\bar{d}$ or $(LLCP)_8\bar{u}d$-mesons into a block of iron two meters thick (the approximate depth of material constituting the central detectors at ATLAS and CMS).  Only charged initial states are considered as LLCPs that hadronize into neutral objects will leave a signal in the calorimeter only and might be difficult to identify. The initial $\beta$ distributions of the LLCPs are taken to be that of $500 \gev$ particles pair produced at the LHC with $\sqrt{s}=10 \tev$, as shown in Fig.~\ref{fig:initialenergy}.  To simulate the effect of the heavy muon trigger \cite{Mermod:2009ct}, we apply a cut of $\beta>0.6$ on this distribution.

\begin{figure}[ht]
\includegraphics[width=0.45\columnwidth]{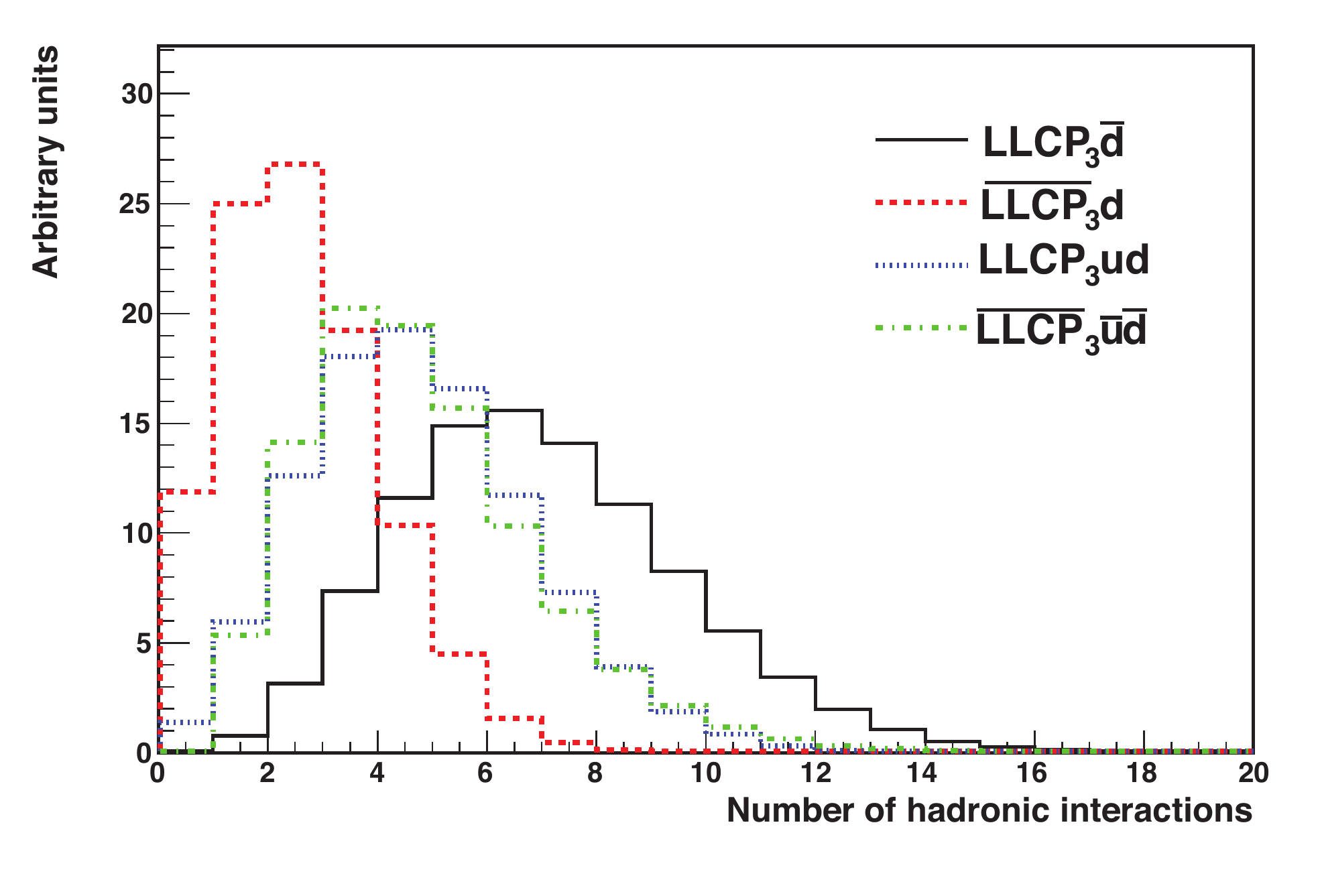}\includegraphics[width=0.45\columnwidth]{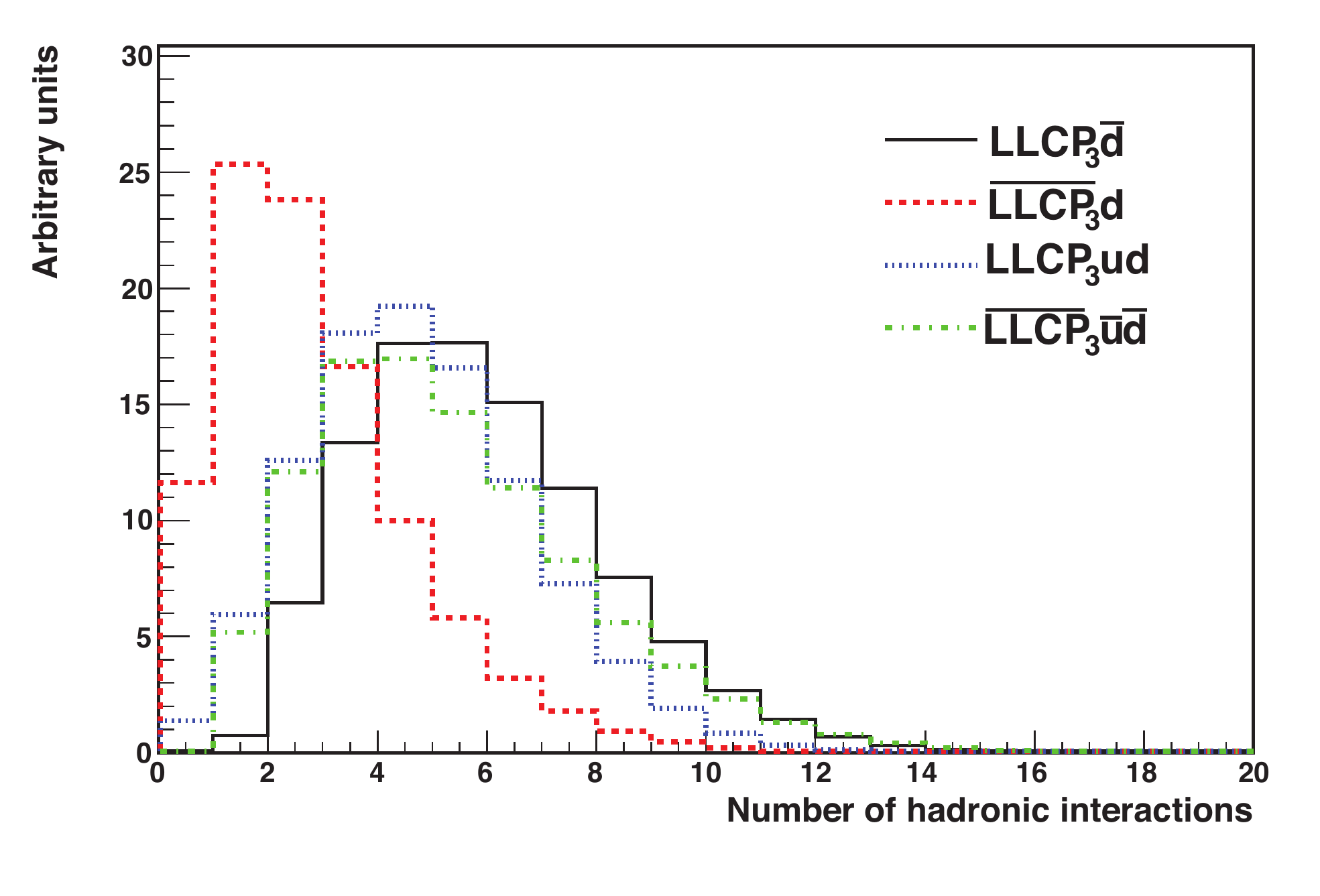}
\includegraphics[width=0.45\columnwidth]{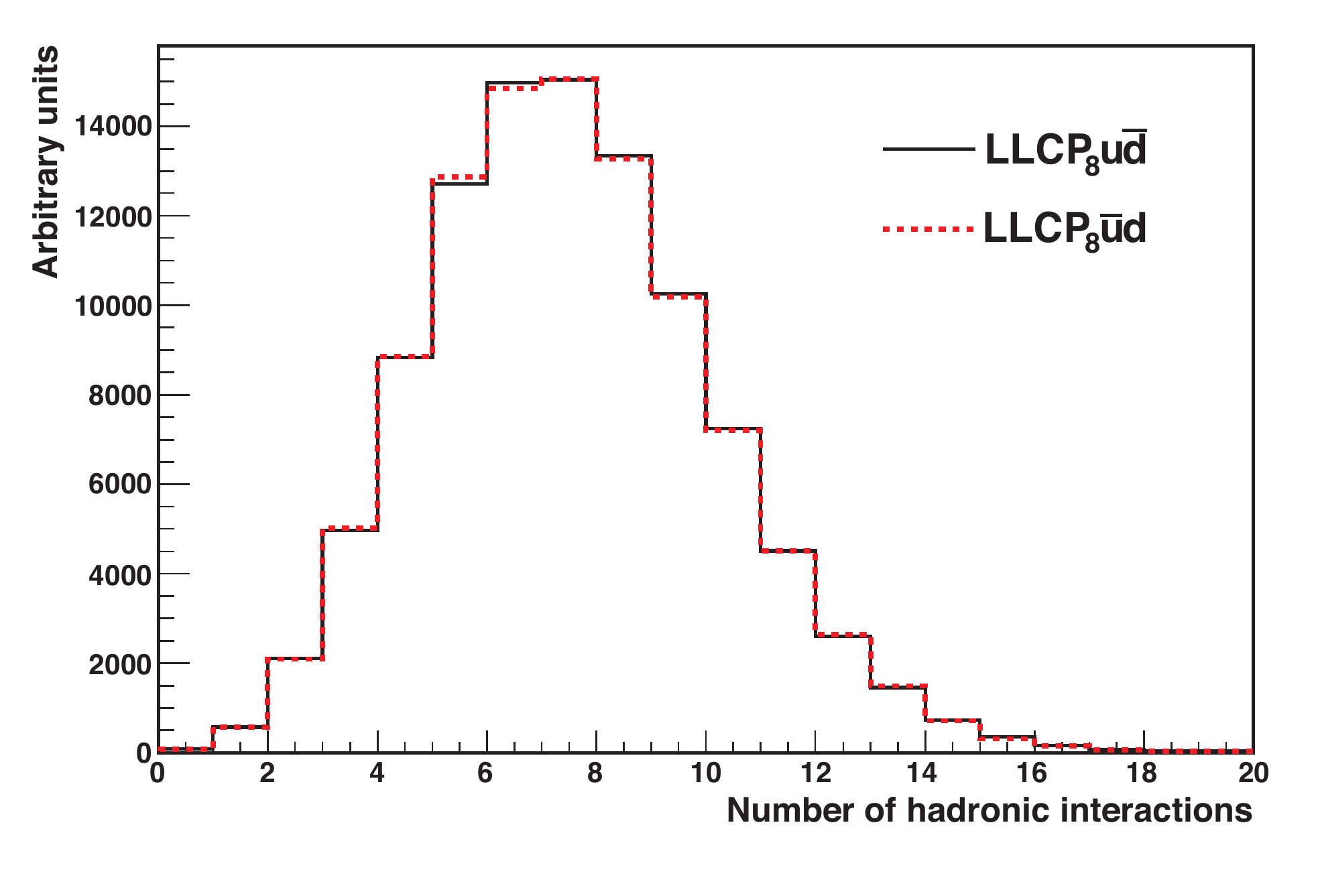}
\caption{The number of hadronic interactions for beams of LLCPs traversing two meters of iron, including the effects of rehadronization after a scattering. The labeling indicates the initial composition of the beam. Top-left panel has no mixing for the neutral LLCP-mesons states, while the top-right panel include maximal mixing. \label{fig:nuclearscattering}}
\end{figure}

The number of nuclear scatterings for different LLCP beams are displayed in Fig.~\ref{fig:nuclearscattering}. As expected, the beam of $\overline{(LLCP)}_3d$-mesons has significantly fewer interactions than the beams of $(LLCP)_3 \bar{d}$-mesons, $(LLCP)_3$-baryons or $(LLCP)_8$-hadrons. Since $\overline{(LLCP)}_3d$ contains an $\overline{(LLCP)}_3$, it cannot rehadronize as a LLCP-baryon, whereas $(LLCP)_3 \bar{d}$ contains $(LLCP)_3$, which tends to rehadonize as a $(LLCP)_3$-baryon with a larger nuclear cross-section. The mixing affects mainly $(LLCP)_3 \bar{d}$-mesons, since they have a larger probability of rehadronizing to a neutral state compared to $\overline{(LLCP)}_3d$, while $(LLCP)_3$-baryons do not undergo significant rehadronization through the detector. Annihilation of $(LLCP)_3$-anti-baryons produces roughly equal amount of charged and neutral $\overline{(LLCP)}_3$-mesons, reducing the sensitivity to mixing. 

\begin{figure}[ht]
\includegraphics[width=0.45\columnwidth]{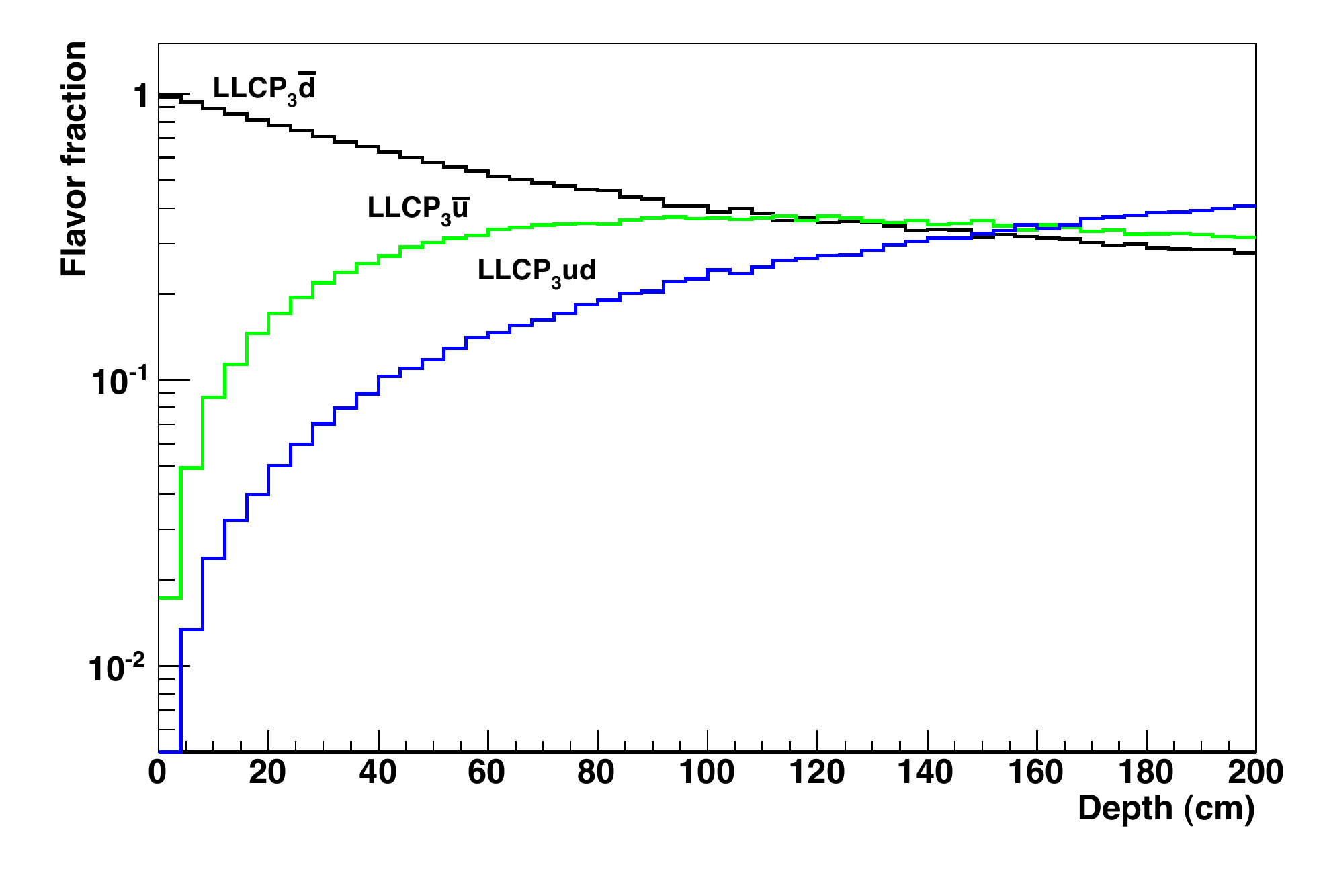}\includegraphics[width=0.45\columnwidth]{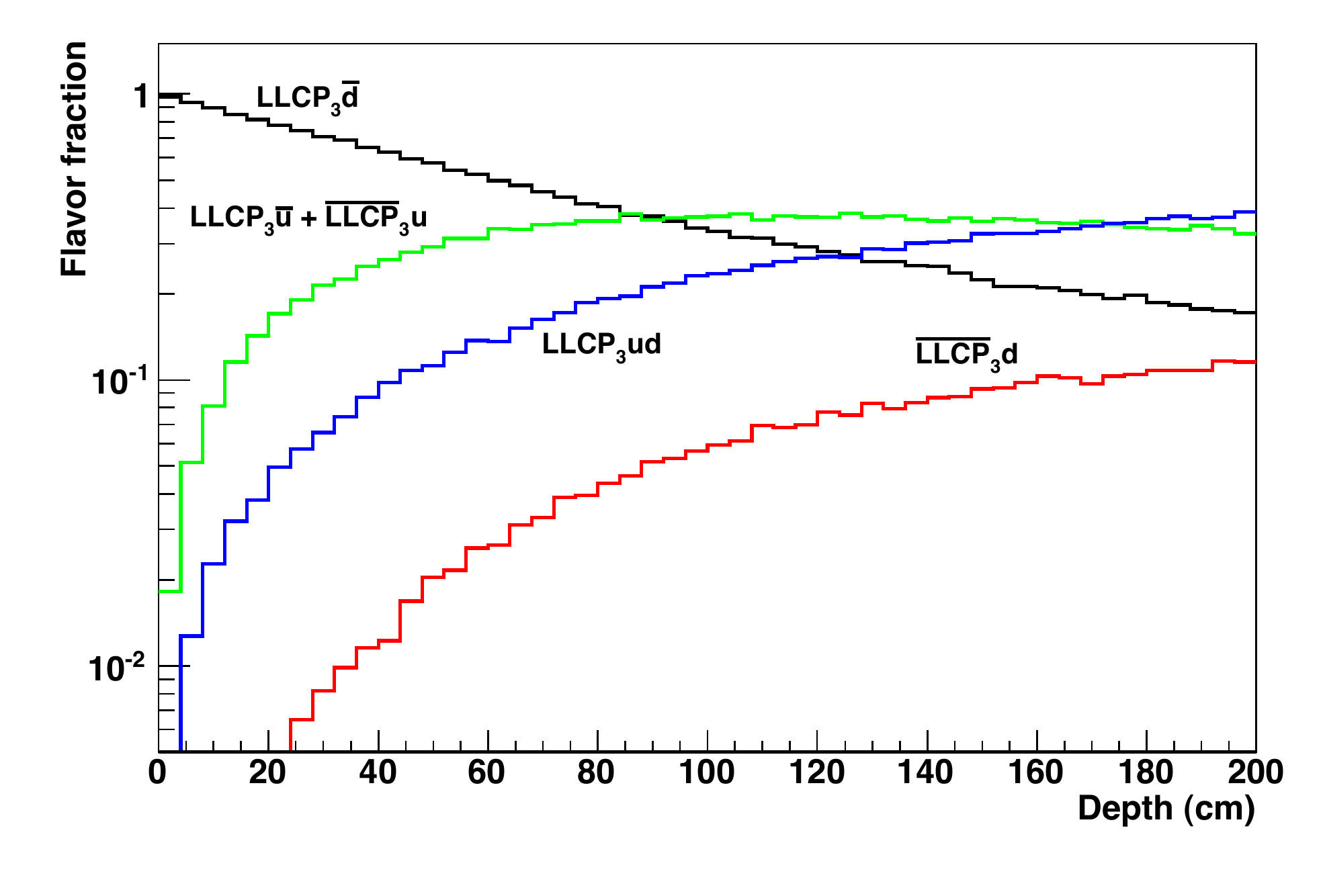}
\includegraphics[width=0.45\columnwidth]{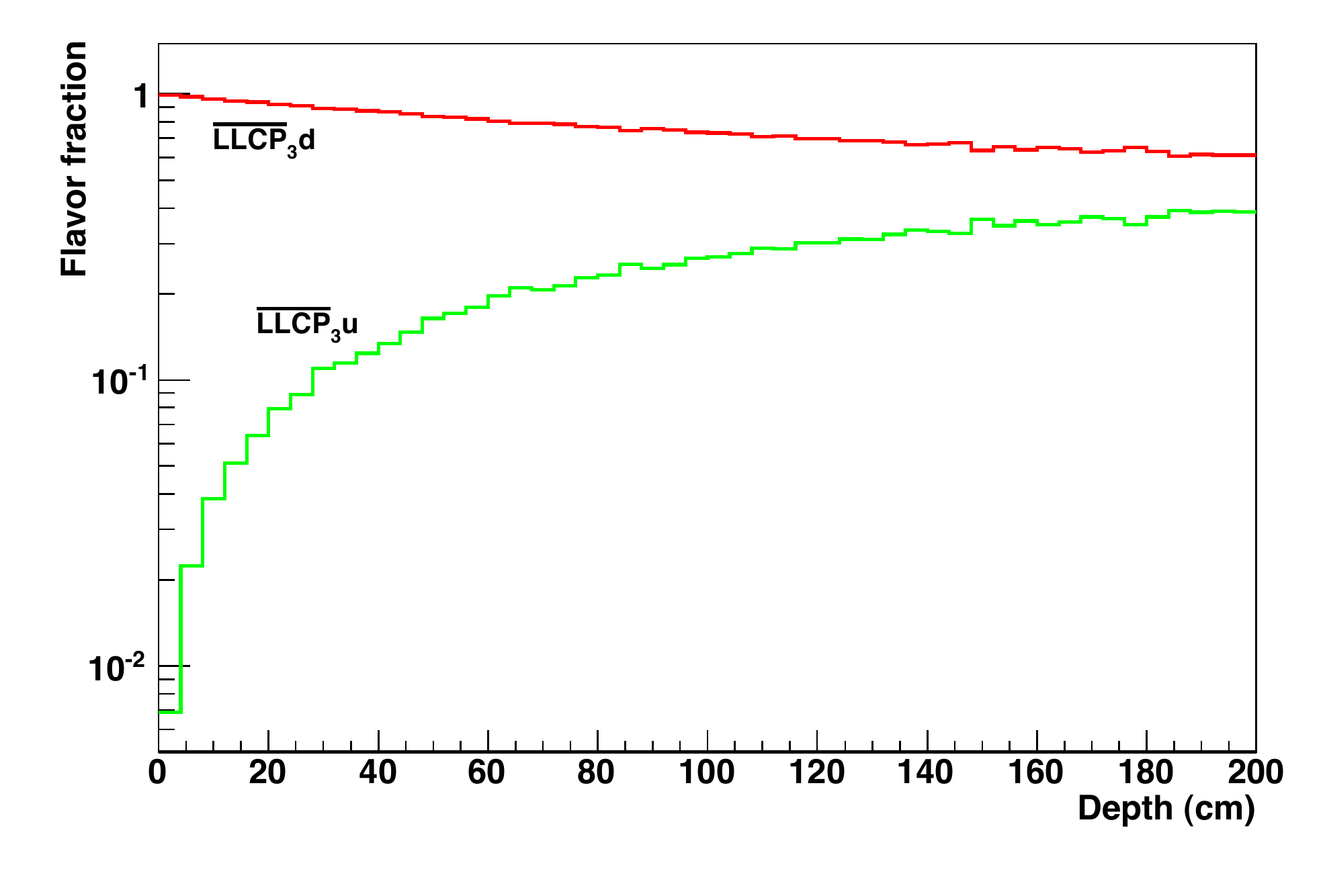}\includegraphics[width=0.45\columnwidth]{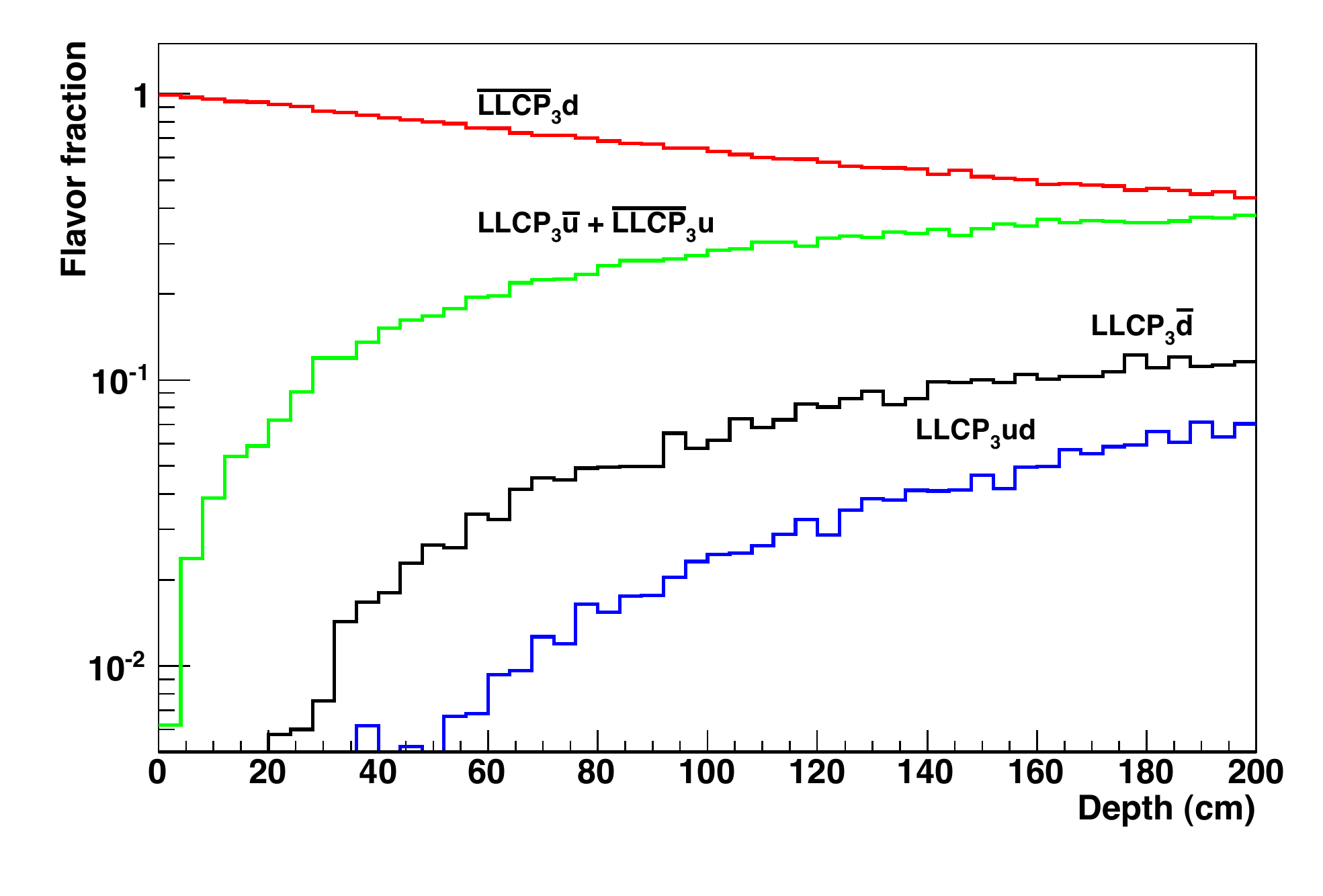}
\includegraphics[width=0.45\columnwidth]{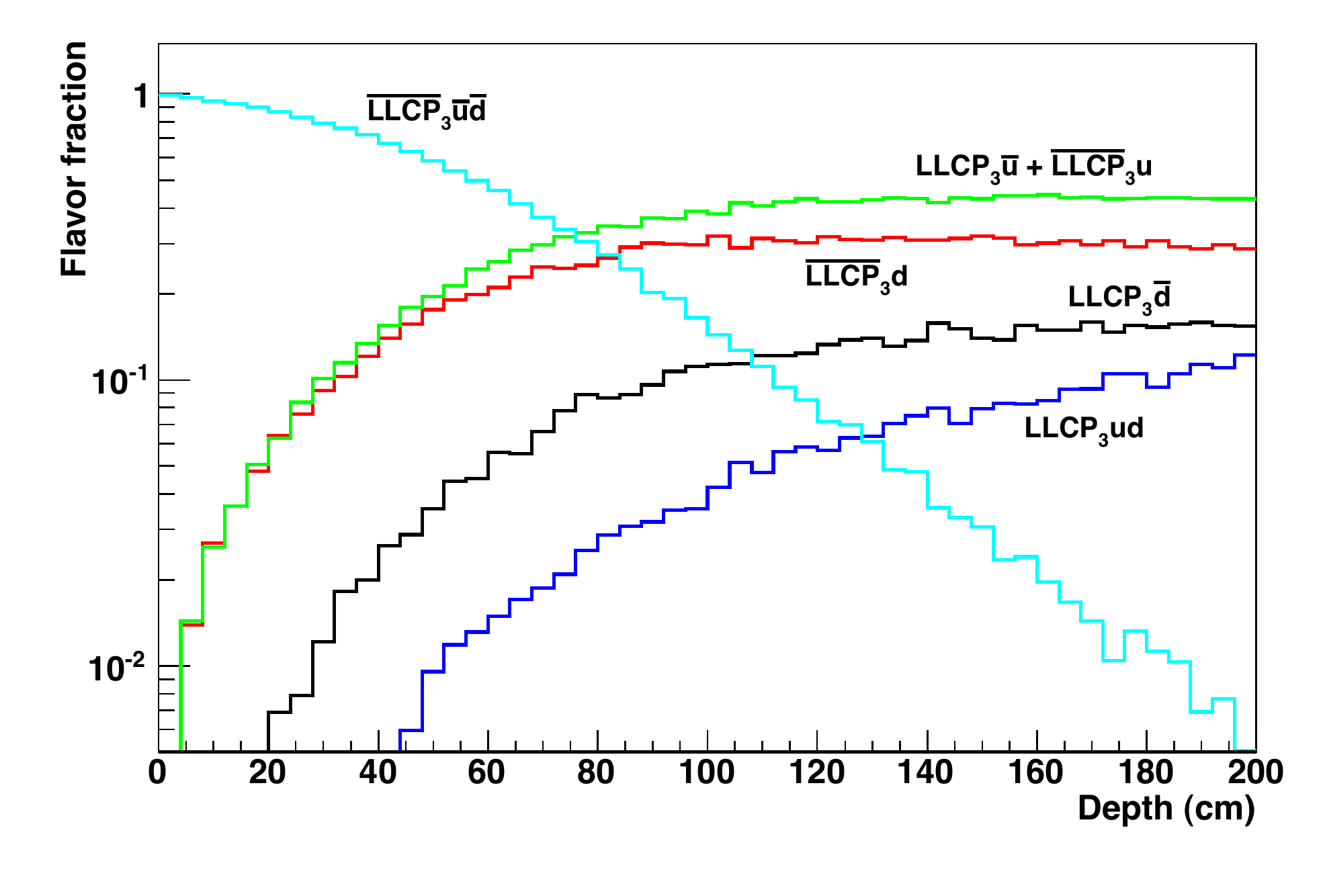}\includegraphics[width=0.45\columnwidth]{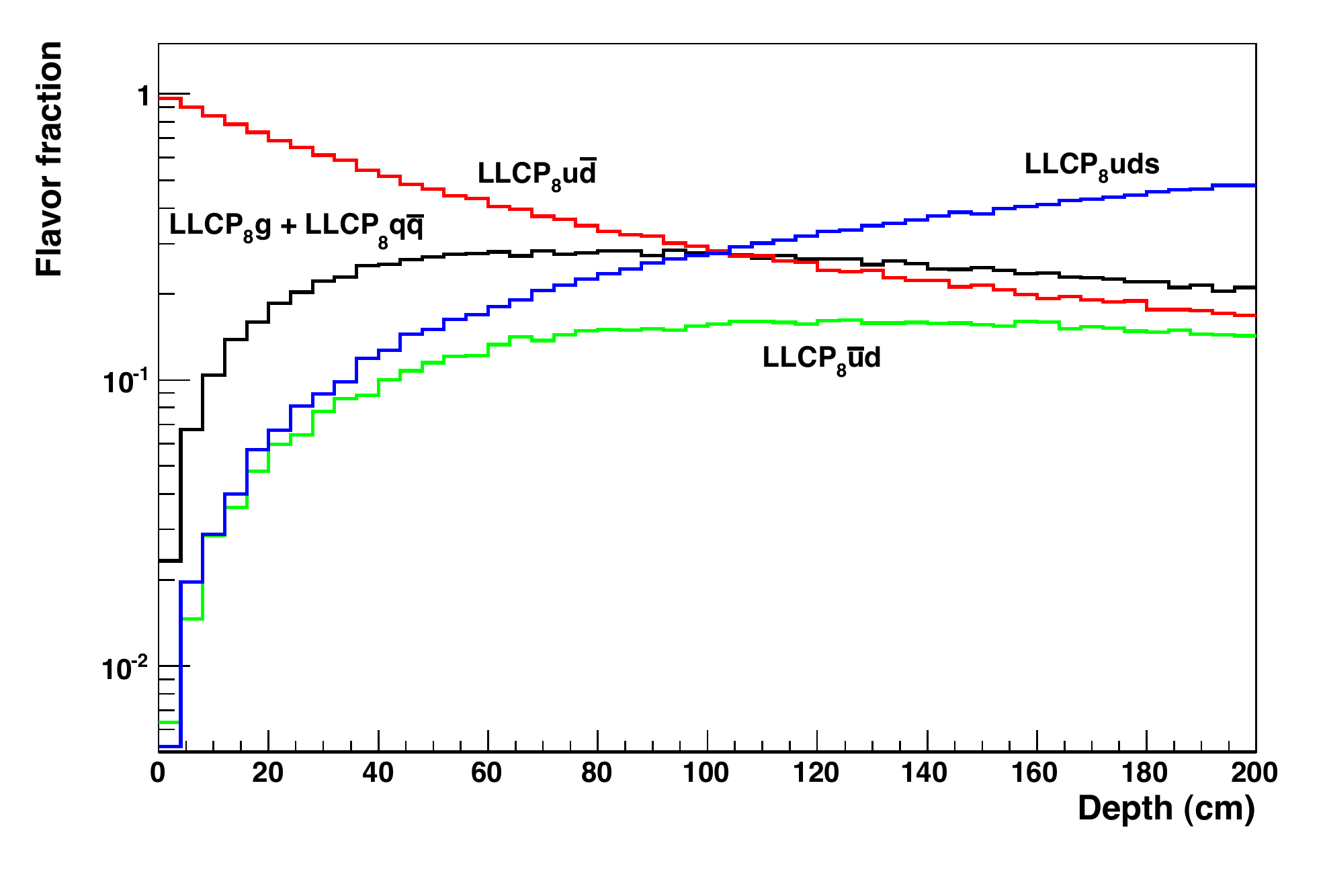}
\caption{The composition of each LLCP beam as a function of distance traveled through iron, assuming an initial beam composed of a pure $(LLCP)_3\bar{d}$ with zero mixing (top left), $(LLCP)_3\bar{d}$ with maximal mixing (top right), $\overline{(LLCP)}_3d$ with zero mixing (middle left), $\overline{(LLCP)}_3d$ with maximal mixing (middle right), $\overline{(LLCP)}_3\bar{u}\bar{d}$-anti-baryons (bottom left), and $(LLCP)_8u\bar{d}$-mesons (bottom right). Note that $(LLCP)_3$-baryons are not shown; as the preferred state for hadronization, they do not undergo significant rehadronization through the detector. \label{fig:rehadronization}}
\end{figure}

In Fig.~\ref{fig:rehadronization}, we show the composition of the beams as they pass through the iron, with and without mixing between $(LLCP)_3$ and $\overline{(LLCP)}_3$. As expected, the beam of $(LLCP)_3$-mesons quickly rehadronizes into baryons, while the $\overline{(LLCP)}_3$-mesons remains stable. Mixing in the neutral meson allows the $\overline{(LLCP)}_3$ beams to develop a small component of LLCP-baryons, but this contribution remains subdominant. We also note that a non-negligible fraction of LLCP-hadrons can undergo charge flips, moving from a positively charged state to a negative one, both in the triplet and octet representations. 

While this provides a signature that is unique to LLCPs, it will certainly complicate track fitting procedures and might be missed in the early running of the LHC. The beam rehadonization simulations indicate that many events will not undergo such sign flips. As these events are not plagued by as many tracking issues, it is these events that we concentrate on in this paper. %reference to Mackeprang et al.

\begin{figure}[ht]
\includegraphics[width=0.45\columnwidth]{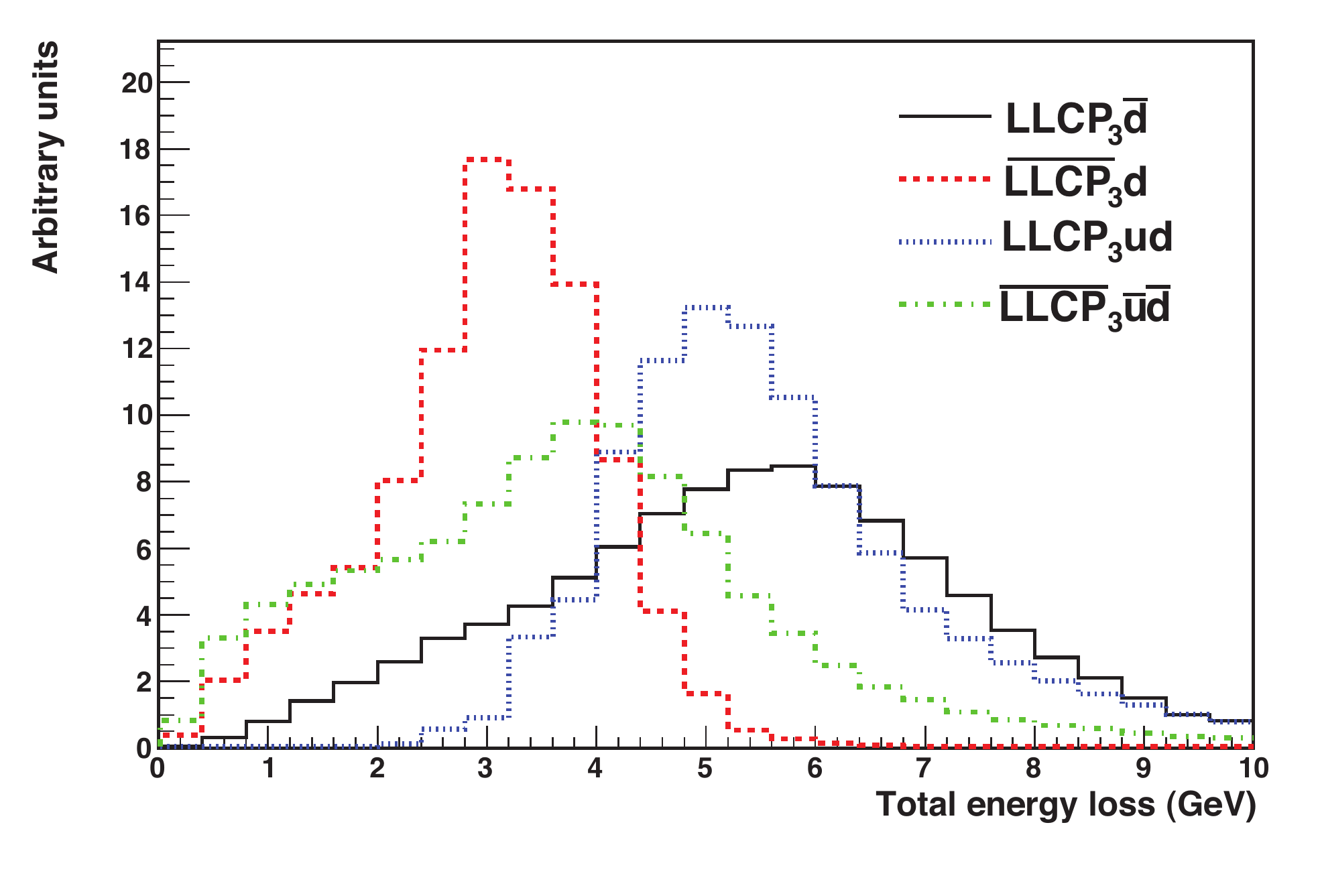}\includegraphics[width=0.45\columnwidth]{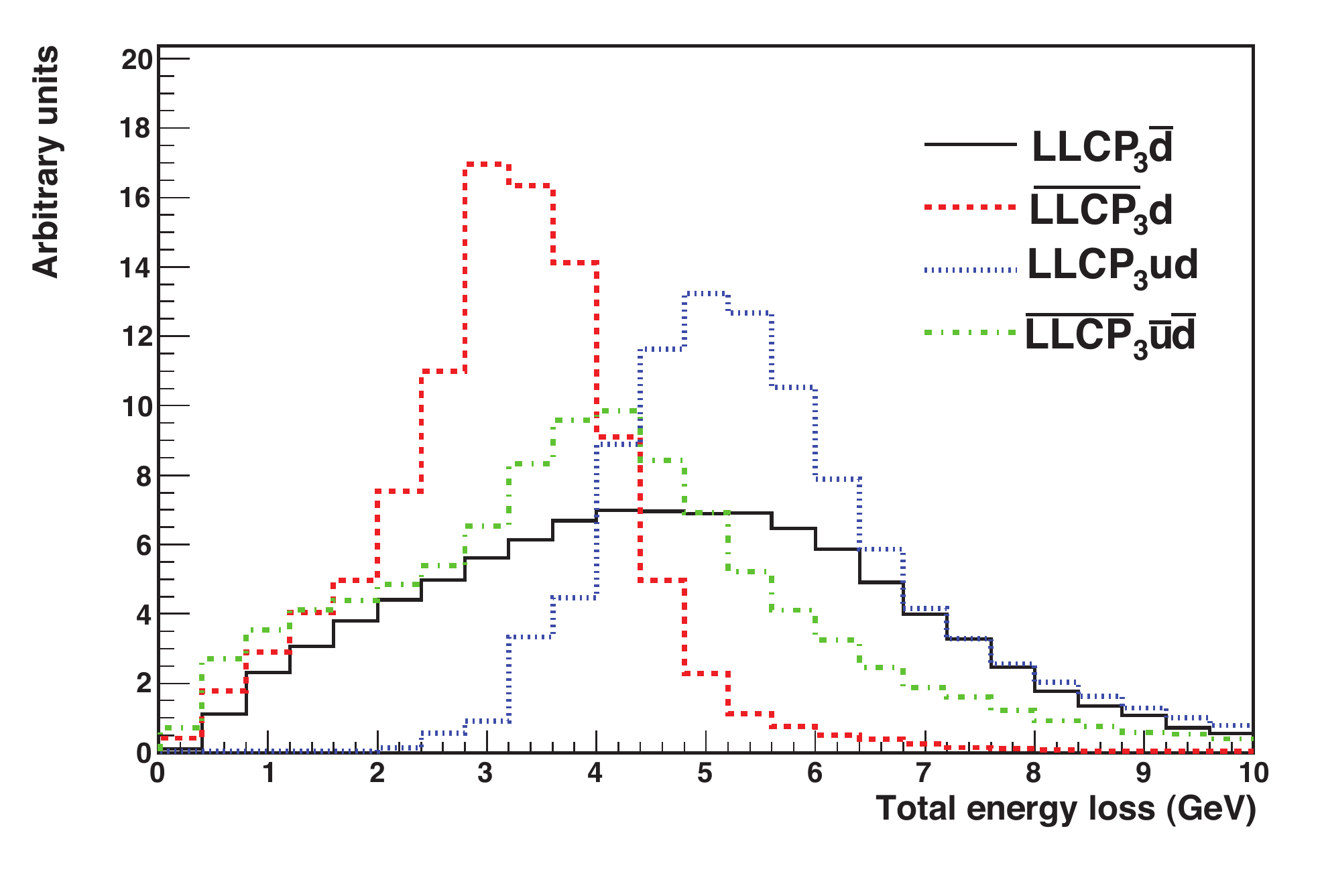}

\includegraphics[width=0.45\columnwidth]{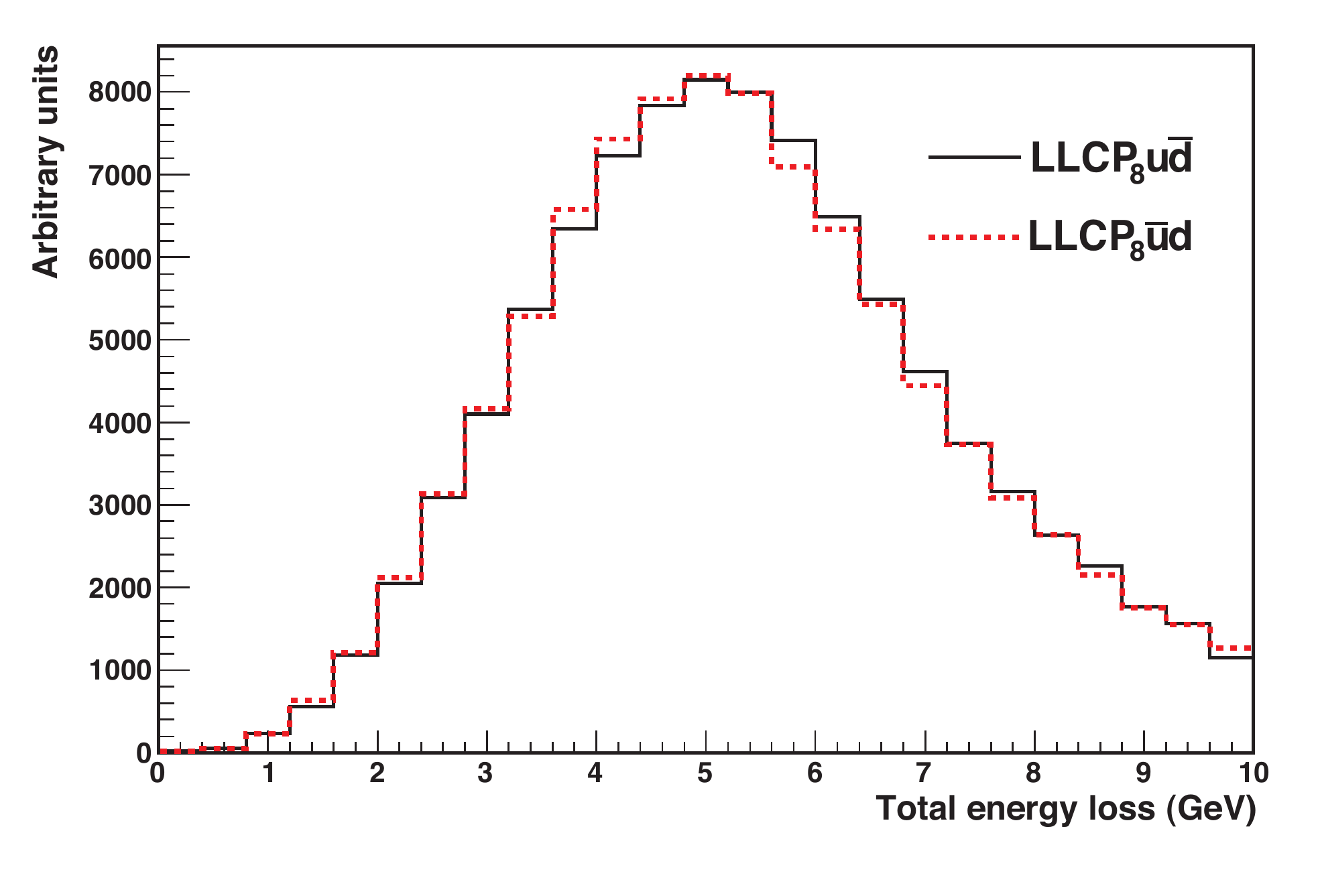}

\caption{Total energy loss through two meters of iron of LLCP beams initially composed of pure $(LLCP)_3$ and $\overline{(LLCP)}_3$ states assuming no mixing (top left) or maximal mixing (top right). The octet states are shown in the lower panel. Labeling indicates the initial composition of each beam. \label{fig:energydep}}
\end{figure}

The total energy deposited in the detector for several charged LLCPs is shown in Fig.~\ref{fig:energydep} and exhibits a similar asymmetry. The $(LLCP)_3$-hadrons leave on average more energy than the $\overline{(LLCP)}_3$-hadrons, regardless of the initial hadronization. As outlined above, the mixing affects mainly $(LLCP)_3 \bar{d}$-mesons, broadening the corresponding distribution. But even with maximal mixing, a significant difference remains. On the other hand, charged $(LLCP)_8$-mesons have similar hadronization schemes and deposit almost equal amount of energy. In pair-production events, the ratio of energy deposited by each track will thus be close to unity, while in triplet/anti-triplet production, a clear asymmetry will be present.

\begin{figure}[ht]
\includegraphics[width=0.45\columnwidth]{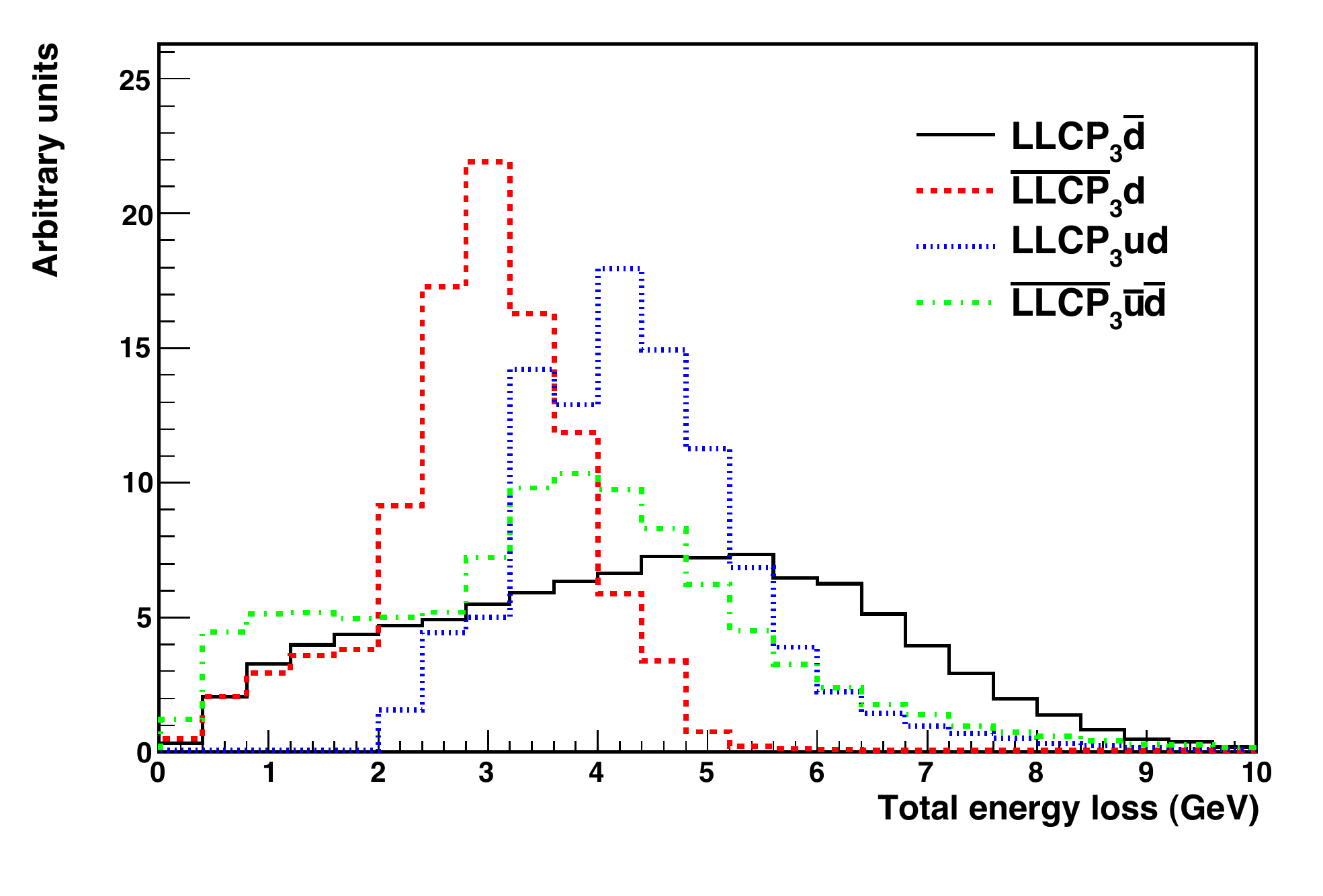}\includegraphics[width=0.45\columnwidth]{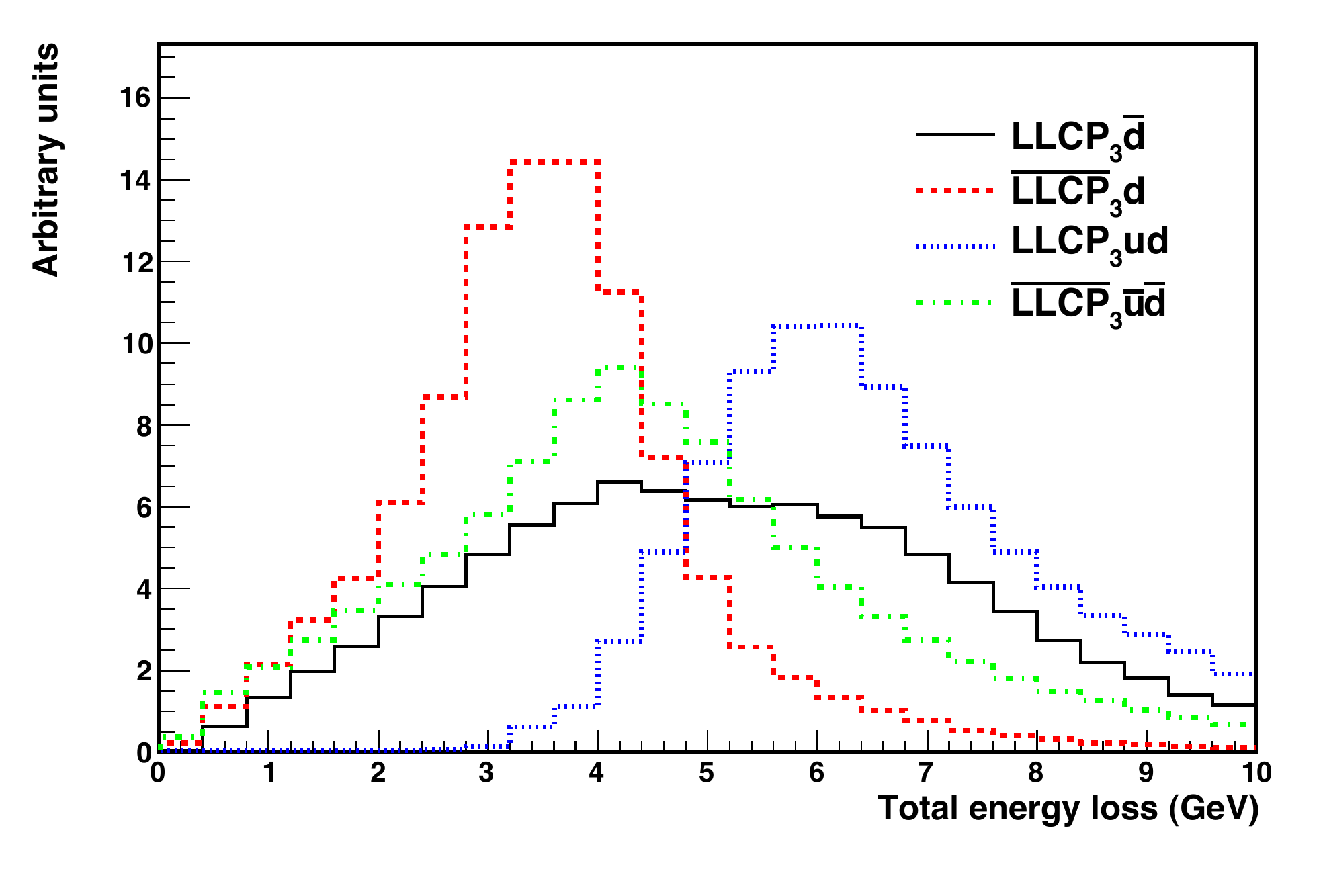}

\caption{Total energy loss through two meters of iron of LLCP beams initially composed of pure $(LLCP)_3$ and $\overline{(LLCP)}_3$ states assuming maximal mixing and a cross section that is $50\%$ larger (left) and $50\%$ smaller (right) than the model implemented in \cite{Milstead:2009qy,Mackeprang:2009ad}.  \label{fig:energydep2}}
\end{figure}

As there is some uncertainty in the hadronic cross section of LLCPs, in Fig.~\ref{fig:energydep2} we plot the total energy loss through two meters of iron of LLCPs with maximal mixing when the hadronic cross section is allowed to vary by $\pm 50\%$. The electromagnetic cross section is held constant. As can be seen, even when the hadronic contribution is decreased by half, the $\overline{(LLCP)_3}$ anti-baryons still deposit considerably less energy than the $(LLCP)_3$-baryons. From this, we conclude that our color charge measurement is robust with regards to uncertainties in the hadronic cross section. 

\section{Conclusions \label{sec:conclusion}}

LLCPs provide an easily recognizable, unique signature at the LHC experiments. The presence of heavy, charged particles with low $\beta$ in the muon chamber is a signal that would be difficult to replicate in the SM. Furthermore, as demonstrated in Fig.~\ref{fig:rehadronization}, rehadronization in the detector can allow charge flips, which would constitute a smoking gun of LLCP-hadron production. We should nonetheless keep in mind that only a fraction of states will undergo charge flips. With large luminosity (varying between $1-1000~\mbox{fb}^{-1}$ depending on the spin of the LLCP), a significant number of events will be accessible by standard track-fitting and analysis techniques. It is these events that we have considered in this analysis of spin and color measurements.

Strongly interacting, stable particles are by no means unique to supersymmetric theories. If discovered, measurements of their fundamental properties: mass, spin, and charge, will be essential to unraveling the degeneracy among possible states. In this paper we demonstrated two experimentally viable measurements to determine the spin and color charge of LLCPs.

To measure spin, we take advantage of the fact that events involving the pair creation of charged LLCPs can be fully reconstructed. As many hadronized states are neutral states, we do not expect every event to contain two visible tracks. However, as seen in Table~\ref{tab:sigma}, the production cross sections are, in most cases, large enough so that hadronization into neutral states should not qualitatively reduce the experimental sensitivity. 

From the two charged tracks, we can reconstruct the center of mass frame of each event, and the polar opening angle of the pair production. In Section~\ref{sec:spin}, we demonstrated that the differential cross section with respect to this angle contains sufficient information to determine the spin of the LLCPs. There is some degeneracy between spin states when heavier intermediaries ({\it i.e.}~gluinos or $KK$ modes) are included. However, in order for these states to significantly affect the measurement, they must be fairly light, and so they should be detectable at the LHC, for example in LLCP plus missing $E_T$ channel.

The measurement of the color charge takes advantage of the rehadronization of the LLCPs inside the detectors. As protons and neutrons contain very few $SU(3)_C$-anti-triplets compared to triplets, there is an asymmetry in how an LLCP in a ${\bf 3}$ representation will rehadronize compared to a $\bar{\bf 3}$. This asymmetry causes the triplet to preferentially hadronize into a baryon, while its anti-partner tends to hadronize into a meson. As the mesons undergo nuclear scatterings less often than the LLCP-baryons, the difference between the ${\bf 3}$ and the $\bar{\bf 3}$ can be experimentally accessed. Using a GEANT4 implementation of the scattering of LLCPs with iron nuclei, we have shown that this asymmetry should be measurable via the energy deposited in the ATLAS and CMS calorimeters. In comparison, octet pairs of LLCPs will not have statistically significant differences between the two tracks, as they will tend to hadronize identically.

It is our expectation that we will be able to distinguish chiral from vector representations, as we have demonstrated with the specific examples of chiral $\bf{3}$ and vector $\bf{8}$ representations. Determining which representation within each set ({\it e.g.}~$\bf{3}$ from $\bf{6}$) will require a more detailed investigation of the energy deposition patterns and a better understanding of the hadronization schemes of representations beyond the adjoint and fundamental. Such a study is beyond the scope of this paper.

\section*{Acknowledgements}
We would like to thank the Aspen Center for Physics for providing a wonderful opportunity for collaboration and discussion. MRB thanks Maria Spiropulu for her advice and support. MRB and BE are supported by the Department of Energy, under grant DE-FG03-92-ER40701 This work is supported in part by the U.S. Dept. of Energy under contact DE-FG02-92-ER40701. LR is supported by NSF grant PHY-0556111. DK is supported by the General Sir John Monash Award.

\bibliography{LLCP}
\bibliographystyle{apsrev}

\end{document}